\documentclass[aps,prl,reprint,superscriptaddress,longbibliography]{revtex4-2}

\usepackage{enumitem}
\usepackage[T1]{fontenc}
\usepackage[utf8]{inputenc}
\usepackage{amsmath,amssymb,amsfonts,braket,physics}
\usepackage[colorlinks=true, linkcolor=blue, citecolor=blue, urlcolor=blue]{hyperref}
\usepackage{bm}
\usepackage{microtype}
\usepackage{mathtools}
\usepackage{comment}

\begin{document}

\title{Continuous Narrow-Linewidth Superradiance in Waveguide QED}
%\title{Continuous Accurate Clock-line Superradiant  Emission in Waveguide QED}
%\title{Continuous Narrow-line Superradiant Emission in Waveguide QED}

\author{Anna Bychek}
\address{Institute for Theoretical Physics,  University of Innsbruck, Technikerstr. 21a, 6020 Innsbruck, Austria}
\affiliation{Institute~for~Quantum~Optics~and~Quantum~Information~of~the~Austrian~Academy~of~Sciences,~6020~Innsbruck,~Austria}
\author{Martin Fasser}
\address{Institute for Theoretical Physics, University of Innsbruck, Technikerstr. 21a, 6020 Innsbruck, Austria}
\author{Ivan Vybornyi}
\address{Institute for Theoretical Physics, Leibniz Universität Hannover, Appelstraße 2, 30167 Hannover, Germany}
%\affiliation{Institute for Quantum Optics and Quantum Information, Austrian Academy of Sciences, 6020 Innsbruck, Austria}
\author{Klemens Hammerer}
\address{Institute for Theoretical Physics, University of Innsbruck, Technikerstr. 21a, 6020 Innsbruck, Austria}
\affiliation{Institute~for~Quantum~Optics~and~Quantum~Information~of~the~Austrian~Academy~of~Sciences,~6020~Innsbruck,~Austria}
\author{Susanne F. Yelin}
\affiliation{Department of Physics, Harvard University, Cambridge, Massachusetts 02138, USA}
\author{Helmut Ritsch}
\address{Institute for Theoretical Physics, University of Innsbruck, Technikerstr. 21a, 6020 Innsbruck, Austria}
\author{Raphael Holzinger}
\address{Institute for Theoretical Physics, University of Innsbruck, Technikerstr. 21a, 6020 Innsbruck, Austria}
\affiliation{Department of Physics, Harvard University, Cambridge, Massachusetts 02138, USA}

\date{\today}

\begin{abstract}
Superradiant lasers promise continuous, narrow-linewidth coherent emission at the bare atomic transition frequency, enabling frequency references of exceptional precision. Recent experiments have advanced the field, but achieving truly continuous operation remains technically challenging. Here we propose an alternative route to an active optical frequency reference with fewer emitters using all-to-all dipole-dipole interactions mediated by a nanophotonic waveguide. We show that selectively pumping only a sub-ensemble of emitters, rather than the full ensemble, substantially improves emission characteristics. The collective interactions with unpumped emitters provide narrowband frequency selection and establish an effective feedback mechanism analogous to the role of a macroscopic cavity. We find directional superradiant emission with strongly phase-synchronized emitter correlations and a narrow output spectrum close to the bare emitter resonance.
Our results demonstrate a strong metrological gain from selective partial pumping of quantum emitters with the second-order intensity correlation $g^{(2)}(0)\simeq 1$, indicating reduced equal-time intensity fluctuations, and open a route to waveguide-based optical frequency references using small clock-atom ensembles for chip-scale precision metrology.
%Our results show a strong metrological gain from selective partial pumping of quantum emitters with the second-order intensity correlation $g^{(2)}(0)\simeq 1$, indicating reduced equal-time intensity fluctuations, and open a route to waveguide-based optical frequency references using small clock-atom ensembles for chip-scale precision metrology.
\end{abstract}

\maketitle

\section{Introduction}
\vspace{-2pt}
Optical atomic clocks achieve exceptional precision by referencing light to long-lived atomic transitions allowing unprecedented accuracy and stability in measuring time and frequency~\cite{oelker2019demonstration,aeppli2024clock}. 
Recently, there has been a growing number of theoretical and experimental studies focused on the development of an \emph{active} optical frequency reference, in which the atoms themselves continuously generate a narrow-linewidth optical signal whose frequency remains tied to the bare atomic transition~\cite{chen2009active,Meiser2009Prospects,Meiser2010SteadyState,Bohnet2012SteadyState,Norcia2016ColdStrontium,Norcia2018FrequencyMeasurements,liu2020rugged,pan2020optical,Kazakov2022UltimateStability,kristensen2023subnatural,fama2024continuous,cline2025continuous,reilly2026fully}.
%However, their short-term stability and deployability remain limited by cavity-stabilized reference lasers in between atomic interrogations~\cite{Ludlow2015OpticalClocks,Grotti2018TransportableClock}.
%but their performance and deployability are strongly influenced by the complexity of trapping, cooling, probing, and reading out large ensembles of atoms~\cite{oelker2019demonstration,aeppli2024clock,Ludlow2015OpticalClocks,Poli2014TransportableSrClock,Grotti2018TransportableClock}.
%An attractive complementary route is an \emph{active} optical frequency reference, in which the atoms themselves continuously generate a narrow-linewidth optical signal whose frequency remains tied to the bare atomic transition~\cite{Meiser2009Prospects,Bohnet2012SteadyState,Norcia2018FrequencyMeasurements}.
Such a source should combine high photon flux, long temporal coherence, and minimal frequency shifts from the clock transition.
%%%%%%%%%%%%%%%%%%%%%%%%%%%%%%%%%%%%%%%%%%%%%
\begin{figure}[ht!]
\centering
\includegraphics[width=0.985\linewidth]{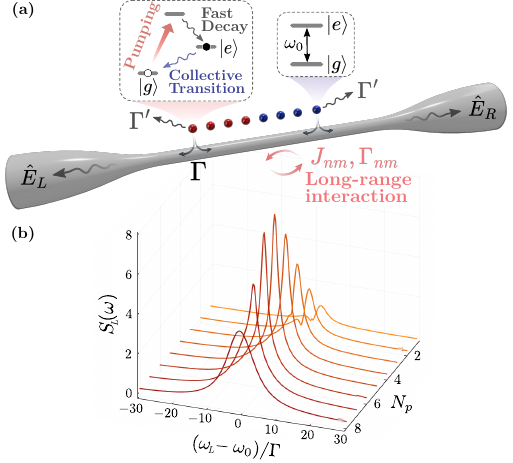}
\vspace{-0.0em}
\caption{\textbf{Continuous narrow-linewidth collective emission.} \textbf{(a)} An ensemble of quantum emitters coupled to a single-mode waveguide reservoir. The emitters feature excited $(|e\rangle)$ and ground $(|g\rangle)$ states with coupling rates $\Gamma$ to the resonant bidirectional waveguide mode and local loss with rates $\Gamma'$ (both assumed to be identical for all emitters). A subset (in red) is incoherently pumped with rate $R$, with the rest (in blue) providing collective feedback, resulting in collective emission into left- and right-propagating electric fields. \textbf{(b)} Emission spectrum from $N=8$ emitters into the left direction, presented for different numbers of pumped emitters $N_p$ at rate $R=5\Gamma$. The spectra show maximal spectral narrowing and peak height for pumped fractions near one-half of the emitters. Simulations are performed with the full solution of the master equation.}
\vspace{-0.0em}
\label{fig1}
\end{figure}
%%%%%%%%%%%%%%%%%%%%%%%%%%%%%%%%%%%%%%%%%%%%%
Collective radiative effects provide a natural mechanism for enhancing useful optical signals. Since Dicke's original work on superradiant spontaneous emission~\cite{Dicke1954,GrossHaroche1982Superradiance}, it has been understood that ensembles of quantum emitters can radiate cooperatively through collective dipole--dipole exchange, producing emission rates and field correlations inaccessible to independent atoms. This makes superradiance a powerful resource for generating macroscopic coherence and collective light emission in driven-dissipative many-body systems~\cite{Lei2023ManyBodyCavityQED,ferioli2021laser,goncalves2025driven,cardenaslopez2026emergentspinordersteadystate,Kersten2026SelfInducedSuperradiantMasing}. For metrological applications, however, the central challenge is not simply to produce a transient superradiant burst, but to harness superradiance as a \emph{continuous narrow-linewidth resource}, where pumping, dissipation, and collective synchronization are balanced in a nonequilibrium steady state without destroying atomic coherence.

A particularly compelling approach to this idea is a \emph{superradiant laser}, where an ensemble of incoherently pumped atomic emitters radiates into a lossy cavity mode in the bad-cavity regime~\cite{Meiser2009Prospects,Meiser2010SteadyState,Bohnet2012SteadyState,Norcia2016ColdStrontium,Norcia2018FrequencyMeasurements,debnath2018lasing,liu2020rugged,pan2020optical,Kazakov2022UltimateStability,kristensen2023subnatural,fama2024continuous,cline2025continuous,reilly2026fully,maier2014superradiant,bychek2021superradiant,schafer2025continuous,Dubey2025ContinuousSuperradiantLaser,hotter2022continuous}. In this regime, optical coherence is stored predominantly in the atoms rather than in the cavity field, reducing the sensitivity of the emitted light to cavity-frequency fluctuations. Theory predicts steady-state superradiant emission with enhanced intensity compared to independent emitters and a spectral linewidth that can be below the natural single-emitter linewidth~\cite{Meiser2009Prospects,Kazakov2022UltimateStability}. These features make superradiant lasers promising candidates for active optical clocks and ultra-stable frequency references. At the same time, recent experimental advances have shown that their development still faces major technical challenges on the route to truly continuous operation, which requires sufficiently large ensembles to be confined, pumped, and replenished while maintaining favorable collective coupling to the optical resonator. Current efforts focus on optical atomic conveyors or beams of excited atoms passing through an optical resonator~\cite{fama2024continuous, schafer2025continuous}, as well as optimized multilevel pumping schemes using multiple repumping lasers~\cite{kristensen2023subnatural,hotter2022continuous}.

This raises a central question for precision metrology: can steady-state superradiant emission be realized in a more compact, cavity-free architecture, with fewer emitters and without relying on a conventional optical resonator? This is directly relevant for frequency metrology and the development of transportable, chip-scale active atomic clocks~\cite{Meiser2009Prospects,Kazakov2022UltimateStability,RileyHowe2008FrequencyStability}. Waveguide QED provides a particularly attractive platform in this context~\cite{RMP_waveguideQED_2023}. When ensembles of quantum emitters couple to a tightly confined one-dimensional guided mode, photons emitted by one emitter can be reabsorbed and reemitted by the others, generating position-dependent coherent emitter--emitter interactions and collective dissipation~\cite{RMP_waveguideQED_2023,Lalumiere2013InputOutput,Mahmoodian2018StronglyCorrelated,Kusmierek2023HigherOrderMeanField}. The relative phases of these interactions can be engineered through the emitter positions along the waveguide, enabling collective radiative states ranging from enhanced \emph{superradiant} to strongly suppressed \emph{subradiant}~\cite{RMP_waveguideQED_2023,cardenas2023many}, while simultaneously collecting the emitted light into directional guided modes~\cite{Kusmierek2025ResonantEnergyTransfer}.

In this work, we show that a {small}, partially pumped ensemble of identical quantum emitters coupled to a bidirectional 1D waveguide realizes a superradiant, narrow-linewidth light source with a single spectral line close to the emitter resonance frequency, as shown in Fig.~\ref{fig1}(b). A key ingredient is \emph{partial incoherent pumping} of the emitter ensemble~\cite{Bychek2025nanoscale,Holzinger2020Nanoscale}. Throughout this work, we use \emph{partial pumping} to denote spatially selective incoherent pumping: only a subset of emitters along the waveguide is optically pumped, while the remaining emitters are left unpumped. % and act as a passive collective resonator.
The pumped fraction provides broadband gain, which is amplified and spectrally filtered by the unpumped emitters. Thus, waveguide-mediated interactions are not only a mechanism for enhanced radiation, but a resource for producing metrologically useful continuous superradiance.

We analyze the steady-state emission properties using the quantum master equation for small system sizes, whereas for larger ensembles we employ a second-order cumulant expansion to capture both the buildup of collective correlations and superradiance as well as the emission spectrum~\cite{Kusmierek2023HigherOrderMeanField}. Photon statistics in the emitted light provide complementary information about equal-time intensity fluctuations, quantified by the zero-delay second-order correlation function $g^{(2)}(0)$~\cite{Ferioli2023NonequilibriumSuperradiant}. This allows us to identify optimal emitter configurations, pump strengths, and pumped fractions that maximize the radiated power and minimize linewidth, and to assess their robustness against static positional disorder and inhomogeneous frequency broadening.

%This paper is organized as follows. We first introduce the bidirectional waveguide master equation, directional output fields, and figures of merit used to characterize the emitted light. We then identify the optimal waveguide phase relation for partially pumped emitter chains. Next, we present the steady-state emission properties, including directional intensity, spectral linewidth, frequency shifts, and photon statistics. We subsequently quantify the metrological performance and assess its robustness against positional and frequency disorder. Finally, we discuss the physical mechanism and outline possible extensions toward reduced collective models and experimental implementations.

\section{Theoretical description}

We consider $N$ \textit{identical} two-level quantum emitters at positions $\{x_n\}$, with resonance frequency $\omega_0$, coupled to a single-mode, bidirectional waveguide reservoir with rate $\Gamma$, and subject to decay into free space outside the resonant waveguide mode with rate $\Gamma'$, as illustrated schematically in Fig.~\ref{fig1}(a). A subset $\mathcal P\subset\{1,\dots,N\}$ consisting of $N_p=|\mathcal P|$ emitters is incoherently pumped with rate $R$, while the remaining emitters are unpumped. After tracing out the electromagnetic field and applying the Born--Markov and rotating-wave approximations, the dynamics of the reduced emitter density matrix $\hat \rho$ are governed by the Lindblad master equation~\cite{Lehmberg1970RadiationI,Agarwal1974QuantumStatistical,Lalumiere2013InputOutput,RMP_waveguideQED_2023}:
\begin{equation}
\dot{\hat{\rho}} = -\frac{i}{\hbar}\big[\hat H,\hat \rho\big] + \mathcal L_{\Gamma}\big[\hat \rho \big] + \mathcal L_\mathrm{R}\big[\hat\rho\big] +\mathcal L_\mathrm{fs}\big[\hat\rho\big],
\label{eq:model_master}
\end{equation}
with coherent exchange described by the Hamiltonian,
\begin{align}
\hat H/\hbar
&= \! \!  \sum_{n,m=1}^N \! \!
J_{nm}\,\hat\sigma_n^\dagger \hat \sigma_m
=
\sum_{n>m}
\frac{J_{nm}}{2}
\big(
  \hat\sigma_n^x \hat\sigma_m^x
 +\hat\sigma_n^y \hat\sigma_m^y
\big),
\label{eq:model_JGamma}
\end{align}
in a frame rotating at frequency $\omega_0$ and with the coherent exchange rate $J_{nm} = \frac{\Gamma}{2} \sin(k |x_n\!-\! x_m|)$. Here, $\hat \sigma_n = |g\rangle_n \langle e|_n$ denotes the Pauli lowering operator of emitter $n$, and using $\hat \sigma_n=\frac{1}{2}(\hat \sigma_n^x-i\hat \sigma_n^y)$, the coherent part forms an $\mathrm{XY}$ spin model without on-site interaction due to $J_{nn}=0$.

Collective dissipation into the waveguide, incoherent pumping and free-space decay are described by the Lindbladian superoperators:
\begin{subequations}\label{eq:lindblad}
\begin{align}
\mathcal L_{\Gamma}[ \hat\rho]
  &=\! \! \sum_{n,m=1}^N \! \frac{\Gamma_{nm}}{2}\Big(2\hat \sigma_n \hat\rho \hat\sigma_m^\dagger
     -  \hat\sigma_n^\dagger \hat \sigma_m\hat\rho  -\hat\rho\hat\sigma_n^\dagger \hat \sigma_m \Big), \label{eq:lindblad_a}\\
\mathcal L_{\text{R}}[ \hat\rho]
  &= \sum_{n \in \mathcal P} \frac{R}{2}\Big(2\hat \sigma_n^\dagger \hat\rho \hat\sigma_n
      -  \hat\sigma_n \hat \sigma_n^\dagger\hat\rho
      - \hat\rho\hat\sigma_n \hat \sigma_n^\dagger \Big),
      \label{eq:lindblad_b}\\
\mathcal L_{\text{fs}}[ \hat\rho]
  &= \sum_{n=1}^N \frac{\Gamma'}{2}\Big(2\hat \sigma_n \hat\rho \hat\sigma_n^\dagger
     -  \hat\sigma_n^\dagger \hat \sigma_n\hat\rho   - \hat\rho\hat\sigma_n^\dagger \hat \sigma_n \Big). \label{eq:lindblad_c}
\end{align}
\end{subequations}
Collective dissipation between emitters $n$ and $m$ occurs with rates $\Gamma_{nm}=\Gamma \cos(k|x_n\!-\!x_m|)$ at the emitter resonance frequency $\omega_0$, where $k=\omega_0/v_g$ is the guided-mode wavevector and $v_g$ is the group velocity of the propagating waveguide photons~\cite{Lalumiere2013InputOutput}. The collective rates are governed by the Green's tensor for a one-dimensional photonic environment evaluated at $\omega_0$ via~\cite{novotny2012principles}
\begin{equation} \label{eq:green}
\frac{3\pi\Gamma}{k} \ \mathbf{d}^* \cdot \mathbf{G}(x_n,x_m,\omega_0) \cdot  \mathbf{d}
= -\frac{i\Gamma}{2} \mathrm{exp}\big({ik|x_n-x_m|}\big),
\end{equation}
where $\mathbf{d}$ is the transition dipole moment of the two-level emitters (using $\mu_0 \omega_0^2 |\mathbf{d}|^2/\hbar=3\pi \Gamma/k$). Furthermore, we assume a bidirectional (non-chiral) waveguide model such that the relative phases between emitters depend on the absolute value of the relative positions, i.e. relative phase $k|x_n-x_m| \ (\text{mod} \ 2\pi)$.
Radiative decay into the waveguide mode competes with free-space losses and results in a coupling efficiency quantified by $\beta \! = \! \Gamma/(\Gamma+\Gamma') \!\leq \! 1$~\cite{RMP_waveguideQED_2023}. We note that, in the superradiant regime, collective emission into the guided mode becomes the dominant mechanism driving the atomic dynamics, surpassing competing dissipative processes.

Although the waveguide-mediated couplings are controlled by the propagation phase $k|x_n-x_m|$, the physical emitter separations can be chosen much larger than the resonant wavelength. 
For an equidistant chain we write the nearest-neighbor spacing as
\begin{equation}
    a=a_0+\delta a,\qquad a_0=\ell\lambda_0,\qquad \ell=0,1,2,\dots,
    \label{eq:physical_spacing}
\end{equation}
so that $ka=2\pi\ell+k\delta a$ and all collective waveguide couplings depend only on the residual phase $\phi = k\delta a$. We therefore assume $a_0\gg \lambda_0$ such that free-space dipole-dipole interactions can be neglected, while $\delta a$ determines the waveguide-mediated coherent and dissipative couplings. As a remark, we note that the collective dissipative couplings can be written in the form
\[
\Gamma_{nm}
=
\Gamma\big[\cos(kx_n)\cos(kx_m)+\sin(kx_n)\sin(kx_m)\big],
\]
as elements of a collective dissipation matrix, which has rank at most two~\cite{cardenas2023many}. Generically, it admits two non-zero eigenvalues
\(\Gamma_\pm\) satisfying \(\Gamma_+ + \Gamma_-=N\Gamma\), while the rank reduces to one in the special case $kx_n \!=\! 2\pi \ell_n$ with $\ell_n \! =0,1,2,\dots$. Physically, this is because a bidirectional 1D waveguide has only two independent radiative output channels: left- and right-propagating modes. Accordingly, \(\mathcal L_\Gamma [\hat{\rho}]\) can be represented by two collective jump operators or, equivalently, by the two directional input-output operators. These representations are related by linear combinations, but the directional output fields are, in general, not identical to the eigenstates of the collective dissipation matrix.

We follow the standard input--output treatment from Refs.~\cite{GardinerCollett1985,Lalumiere2013InputOutput} for the left- and right-propagating electromagnetic fields, $\hat{a}_{L/R}(t) \!=\!\hat{a}_{L/R,\mathrm{in}}(t) + \sum_j e^{\pm ikx_j}\sqrt{\Gamma/2}\ \hat{\sigma}_j(t)$~\cite{Lalumiere2013InputOutput}. For a single guided mode in units of the group velocity $v_g$ (set to unity below), the positive-frequency components of the fields in the far field to the right/left of the emitter ensemble read
\begin{equation}
\hat E_{L/R} = i \sqrt{{\Gamma}/{2}}
\sum_{n=1}^N e^{\pm i k x_n} \hat\sigma_n .
\label{eq:ERL_atomic}
\end{equation}
Here, we have omitted the input fields $\hat E_{\mathrm{in}}$ appearing in the input--output relations. In our setup the guided modes are assumed to be initially in the vacuum state, and we evaluate normally ordered output-field observables, for which the vacuum input contributes only quantum noise and does not affect the quantities considered below. We therefore use Eq.~\eqref{eq:ERL_atomic} throughout this work as the radiated field exiting the waveguide through the left- and right-propagating output fields.

\subsection{Figures of merit}

%%%%%%%%%%%%%%%%%%%%%%%%%%%%%%%%%%%%%%%%%%%%%%%%%%%%%%%%%%%%
%%%%%%%%%%%%%%%%%%%%%%%%%%%%%%%%%%%%%%%%%%%%%%%%%%%%%%%%%%%%
\begin{figure}[ht!]
  \centering

  \makebox[\columnwidth][c]{%
    \hspace*{-0.06\columnwidth}%
    \includegraphics[width=0.81\columnwidth]{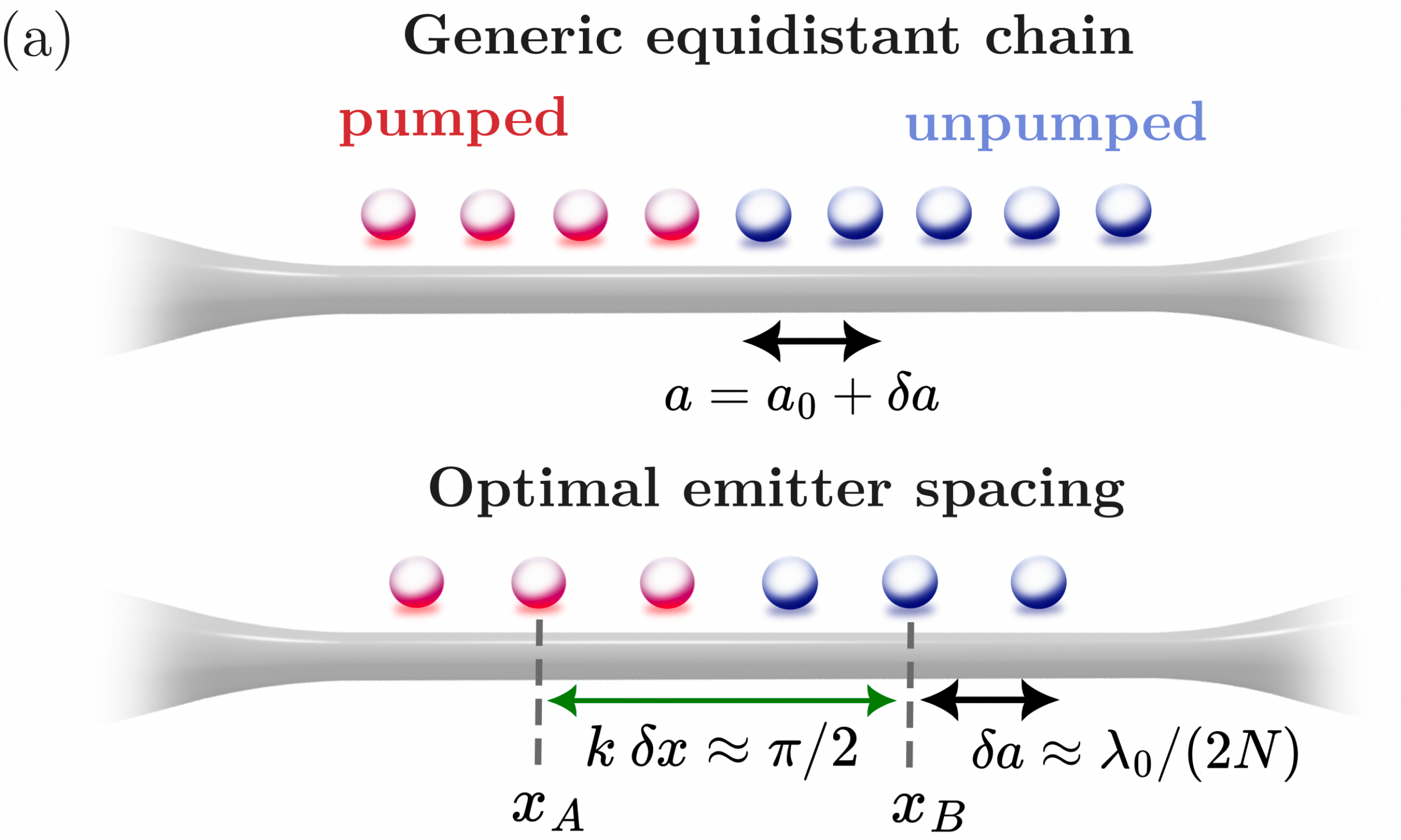}%
  }

  \includegraphics[width=0.99\columnwidth]{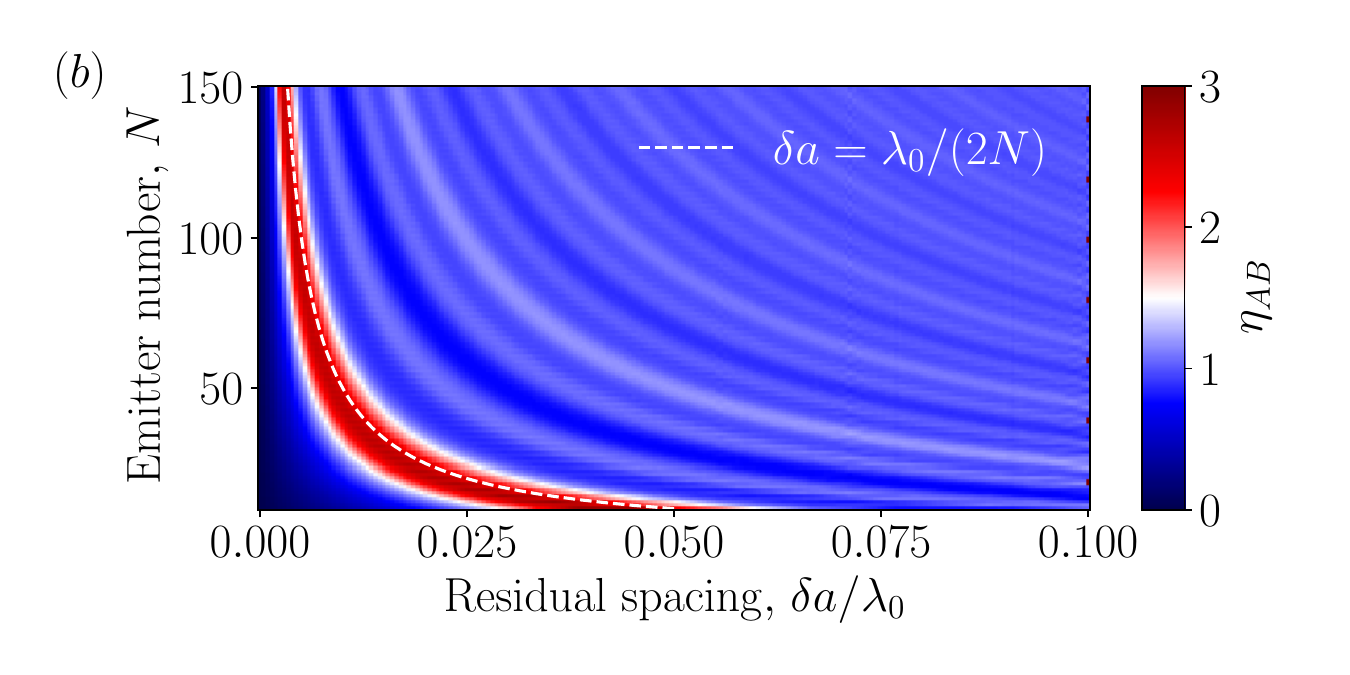}

  \includegraphics[width=0.93\columnwidth]{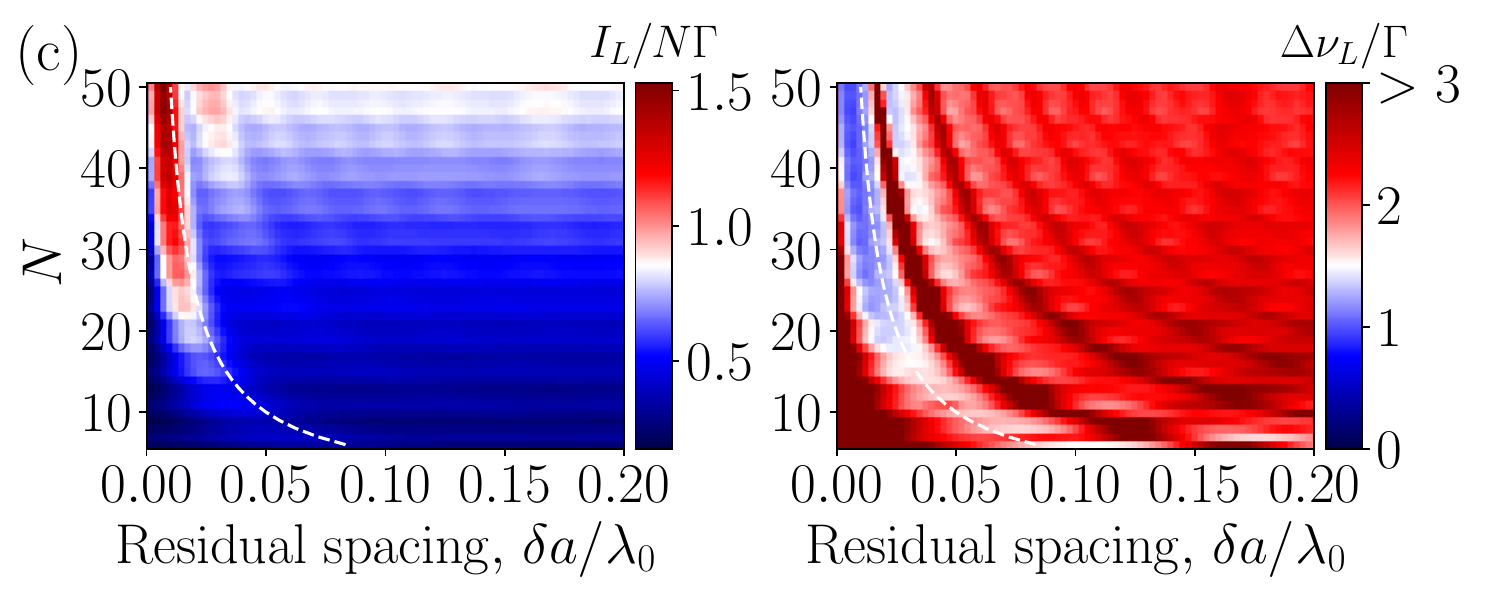}

  \caption{\textbf{Optimal emitter spacing and emission properties of a regular chain under incoherent pumping of its left half:} \textbf{(a)} Sketch of a chain with nearest-neighbor emitter spacing $a=a_0+\delta a$, where $a_0=\ell\lambda_0$ is an integer-wavelength offset. For an optimized emitter spacing, center-to-center residual phase is approximately $k \delta x \approx \pi/2$ between the two sub-ensemble centers. \textbf{(b)} The ratio of coherent versus dissipative coupling between the pumped (A) and unpumped (B) ensembles is quantified by the parameter $\eta_{AB}$ defined in Eq.~(\ref{eq:eta}) and plotted as a function of the residual spacing $\delta a$ and total emitter number $N$. For $N_p=N/2$, the optimal residual spacing follows $\delta a\approx \lambda_0/(2N)$ (white dashed line) corresponding to a residual phase $k\delta a\approx \pi/N$. \textbf{(c)} The numerical results for the steady-state emission intensity ($I_L$) and spectral linewidth ($\Delta \nu_L$) exhibit optimal values when $\eta_{AB}$ is maximal (white dashed line). Simulations are based on the second-order cumulant approach. Parameters: $R=10\Gamma, \Gamma'=0$.}
  \label{fig2}
\end{figure}
%%%%%%%%%%%%%%%%%%%%%%%%%%%%%%%%%%%%%%%%%%%%%%%%%%%%%%%%%%%%

We characterize the steady-state emission properties by analyzing the following:

\paragraph{(i) Emitted field intensity.}
A central signature of steady-state superradiance is a superlinear (up to quadratic) scaling of the emitted intensity with the emitter number $N$~\cite{Dicke1954,ferioli2021laser}. The radiated steady-state field intensity (or power divided by $\hbar\omega_0$) into the waveguide reads
\begin{equation}
{I}_\mathrm{L/R} \! = \! \big\langle \hat E_\mathrm{L/R}^{\dagger}\hat E_\mathrm{L/R} \big\rangle \!=\! \frac{\Gamma}{2} \! \sum_{n,m=1}^N e^{\pm i k (x_m-x_n)} \langle \hat \sigma_n^\dagger \hat \sigma_m \rangle.
\label{eq:emission}
\end{equation}
Independent emission yields $I_\mathrm{L/R} \propto N\Gamma$, whereas phase-synchronized emission leads to superlinear scaling of the emitted intensity with $N$.

\paragraph{(ii) Spectral properties of the emitted light.}
A further figure of merit is a narrow, well-resolved emission line with full width at half maximum (FWHM) %$\Delta \nu \sim \Gamma$, 
centered close to the bare emitter frequency $\omega_0$. The normally ordered, steady-state emission spectra in the left/right directions are defined by
\begin{equation}
S_\mathrm{L/R}(\omega) = 2\,\mathrm{Re}\! \left[
 \int_{0}^{\infty}\! d\tau\, e^{\mathrm{i} \omega \tau}\,
\big\langle \hat E_\mathrm{L/R}^{\dagger}(t_\mathrm{ss}{+}\tau)\, \hat E_\mathrm{L/R}(t_\mathrm{ss})\big\rangle \right],
\label{eq:spectrum}
\end{equation}
where $t_\mathrm{ss}$ denotes the simulation time at which the system is observed to have reached steady state.

\paragraph{(iii) Photon statistics.}
Finally, we characterize the photon statistics of the emitted field via the equal-time, zero-delay intensity correlation function
\begin{equation}
g^{(2)}_\mathrm{L/R}(0) = 
\frac{\Big\langle 
\big(\hat E_\mathrm{L/R}^\dagger (t_\mathrm{ss}) \big)^2
\big(\hat E_\mathrm{L/R}(t_\mathrm{ss}) \big)^2
\Big\rangle}{\Big\langle 
\hat E_\mathrm{L/R}^\dagger (t_\mathrm{ss}) 
\hat E_\mathrm{L/R} (t_\mathrm{ss}) 
\Big \rangle^{2}},
\label{eq:g2-function}
\end{equation}
which probes equal-time intensity fluctuations of the left- and right-going output fields. Values $g^{(2)}_\mathrm{L/R}(0)\!=\!1$ correspond to Poissonian photon statistics, $g^{(2)}_\mathrm{L/R}(0)\!>\!1$ to photon bunching (super-Poissonian statistics), and $g^{(2)}_\mathrm{L/R}(0)\!<\!1$ to photon antibunching (sub-Poissonian statistics). In our system, $g^{(2)}_\mathrm{L/R}(0)$ complements the intensity scaling and linewidth by quantifying the strength of equal-time intensity fluctuations in the directional output fields. 

\subsection{Optimal atomic emitter spacing}

In physical implementations, the emitters do not need to be separated by subwavelength distances, since the coherent and dissipative waveguide-mediated couplings depend only on the residual phase $\phi=k\delta a$. In this work, we divide the chain into a pumped ensemble $A=\{1,\dots,N_p\}$ followed by an unpumped ensemble $B=\{ N_p  +1,\dots,N\}$.
To characterize how the passive ensemble interacts with a given pumped emitter $n\in A$, we introduce the local phase sum
\begin{equation}
r_n= \sum_{j\in B} \mathrm{exp}\big[{ik(x_j-x_n)}\big]
=\! \!\!\!\! \sum_{j=N_p+1}^{N} \!\!\!\mathrm{exp}\big[{i\phi(j-n)}\big] ,
\end{equation}
where the integer-wavelength part of the spacing drops out of the phase factor. For an equidistant chain, this geometric series can be evaluated in closed form as
\begin{equation}
r_n = \mathrm{exp} \left[i\phi\left(\frac{3N}{4}+\frac12-n\right)\right]
\frac{\sin(N\phi/4)}{\sin(\phi/2)}.
\label{eq:rn_halfpumped}
\end{equation}
The prefactor in Eq.~(\ref{eq:rn_halfpumped}) is the propagation phase from emitter $n$ to the center of the passive ensemble, while the common amplitude factor describes the coherent addition of all contributions from ensemble $B$. The relative propagation phase therefore controls whether the inter-ensemble coupling is predominantly coherent or dissipative.

To quantify this, we define the accumulated coherent and dissipative inter-ensemble contributions and their ratio as
\begin{equation}
\label{eq:eta}
J_{AB} \!=\! \sum_{n\in A} \big|\Im r_n\big|,
\
\Gamma_{AB} \!= \!\sum_{n\in A} \big|\Re r_n\big|,
\quad
\eta_{AB}\equiv \frac{J_{AB}}{\Gamma_{AB}}.
\end{equation}
The ratio $\eta_{AB}$ is a measure for how strongly coherent exchange dominates over dissipative coupling between the pumped and passive halves. 

In Fig.~\ref{fig1}(b), we observe that the improved emission characteristics are obtained when the pumped and unpumped sub-ensembles are of comparable size.
In the following, we focus on the representative case where the first half of the emitters is incoherently pumped, while the second half remains unpumped (more general pumped fractions are discussed in the Supplemental Material). 
In the half-pumped case, the common amplitude factor in Eq.~\eqref{eq:rn_halfpumped} cancels in $\eta_{AB}$, so that the optimization is governed purely by the phase profile across the pumped ensemble.

The analytical results presented in Fig.~\ref{fig2}(b) indicate that $\eta_{AB}$ is maximized close to
\begin{equation}
    \phi_{\mathrm{opt}}=k\delta a_{\mathrm{opt}}\approx \frac{\pi}{N}.
\end{equation}
For $k=2\pi/\lambda_0$, this corresponds to a residual spacing $\delta a_{\mathrm{opt}}\approx {\lambda_0}/({2N})$, and therefore to an actual emitter spacing
\begin{equation}
    a_{\mathrm{opt}}=a_0+\delta a_{\mathrm{opt}}
    =
    \ell\lambda_0+\frac{\lambda_0}{2N}.
\end{equation}
 This agrees well with the residual spacing at which the steady-state directional emission is maximal and the spectral linewidth is minimal, shown in Fig.~\ref{fig2}(c).
 %as calculated numerically in Fig.~\ref{fig2}(c) and described in the next section.

This optimum admits a simple geometric interpretation in terms of the phase difference between the centers of the two ensembles. The center positions of the pumped and passive halves are
\begin{equation}
x_A = a\left(\frac{N}{4}+\frac12\right),
\qquad
x_B = a\left(\frac{3N}{4}+\frac12\right),
\end{equation}
so that their center-to-center separation is $x_B-x_A=Na/2$. Since the integer-wavelength part contributes only an integer multiple of $2\pi$ to the inter-emitter phases, the relevant reduced center-to-center phase is $k(x_B-x_A) \equiv {N\phi}/{2}\  (\mathrm{mod}\ \pi)$. At the optimal residual phase $\phi_{\mathrm{opt}}=\pi/N$, this becomes $k(x_B-x_A)=\pi/2 \  (\mathrm{mod}\ \pi)$. Thus, the optimal spacing corresponds to an approximate quarter-wave phase shift between the centers of the pumped and passive halves, which suppresses the net dissipative inter-ensemble coupling while enhancing the coherent exchange responsible for collective feedback and narrow-linewidth superradiant emission.

\section{Steady-state results}

%%%%%%%%%%%%%%%%%%%%%%%%%%%%%%%%%%%%%%%%%%%%%%%%%%%%%%%%%%%%%%%%%%%%%%%%%%%%%%%%%%%%%%%%%%
\begin{figure}[t]
  \centering
  \includegraphics[width=0.9\columnwidth]{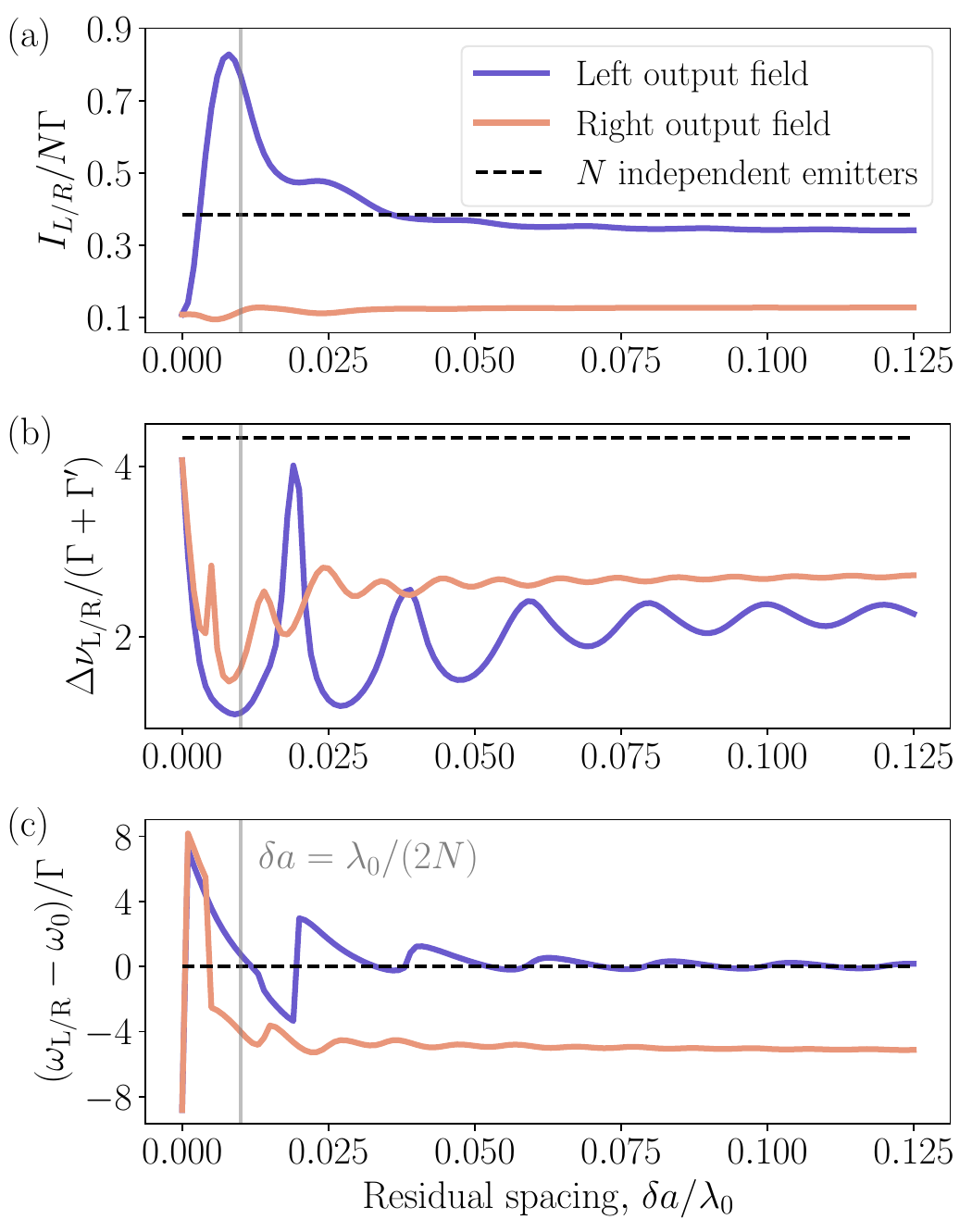}
 \caption{\textbf{Emission into left- and right-propagating waveguide modes.} Emerging left/right asymmetry of steady-state emission properties for varying residual nearest-neighbor phase $k\delta a$. \textbf{(a)} Normalized steady-state photon emission $I_{L/R}$ is enhanced into the left and strongly suppressed into the right-propagating direction. \textbf{(b)} Full width at half maximum $\Delta \nu_{L/R}$ of the spectral peak in units of the emitter spectral linewidth $\Gamma + \Gamma'$ decreases strongly toward the left output at optimal spacing. \textbf{(c)} The spectral peak position $\omega_{L/R}$ is close to the emitter's natural frequency for left propagation and strongly shifted for right propagation. Parameters: $N_p=N/2$, $N=50$, $R=10\Gamma$, $\Gamma'=2\Gamma$.}
  \label{fig3}
\end{figure}
%%%%%%%%%%%%%%%%%%%%%%%%%%%%%%%%%%%%%%%%%%%%%%%%%%%%%%%%%%%%%%%%%%%%%%%%%%%%%%%%%%%%%%%%%%

We now turn to the steady-state emission properties of a partially pumped emitter chain coupled to a single-mode waveguide reservoir. 
In the following, we calculate the dynamics and steady-state properties using a second-order cumulant expansion approach~\cite{kubo1962generalized, rubies2023characterizing} that retains two-body correlations, see the Supplemental Material for details. In related physical models, this approach has captured the steady-state emission properties quite well and shown good agreement with the exact solution of the master equation~\cite{Bychek2025nanoscale,rubies2023characterizing}.
%The steady state is obtained from the cumulant equations of motion (see Supplemental Material), and the directional spectra are evaluated from the corresponding two-time correlation functions. 
The central observables are the directional photon emission $I_{L/R}$, the spectral linewidth $\Delta\nu_{L/R}$, the frequency shift $\omega_{L/R}-\omega_0$ of the spectral peak from the emitter resonance, and the equal-time intensity correlation $g^{(2)}_{L/R}(0)$.

Figure~\ref{fig2} and the ratio in Eq.~({\ref{eq:eta}}) first identify the geometric condition under which the pumped and unpumped parts of the chain are predominantly coupled via coherent interaction. At the optimal residual spacing $\delta a_{\rm opt}\simeq \lambda_0/(2N)$ 
%, the center-to-center phase between the two halves is $\pi/2$ modulo $\pi$, so that 
the net dissipative coupling between the two ensembles is suppressed while the coherent exchange remains large. The steady-state emission and spectral linewidth properties in Fig.~\ref{fig2}(c) show that this geometric optimum coincides with enhanced directional emission and a reduced spectral linewidth. %This indicates that the passive half of the chain acts as a phase-sensitive collective resonator for the incoherently pumped atoms.

%%%%%%%%%%%%%%%%%%%%%%%%%%%%%%%%%%%%%%%%%%%%%%%%%%%%%%%%%%%%%%%%%%%%%%%%%%%%%%%%%%%%%%%%%%
\begin{figure}[t]
  \centering
  \includegraphics[width=0.85\columnwidth]{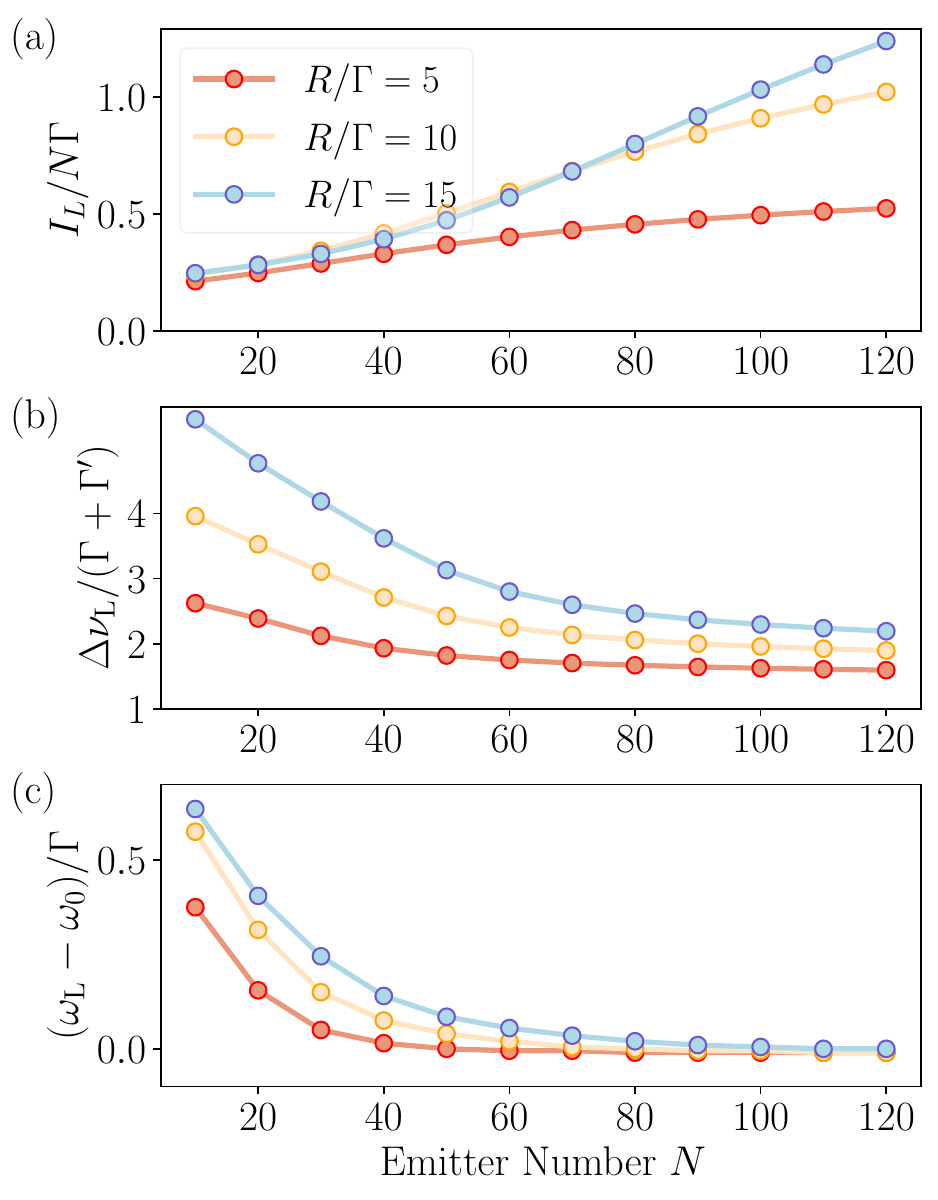}
 \caption{\textbf{Equidistant emitter chain.}
 The emission properties are shown for the left-propagating field as a function of the total emitter number $N$ for three different pumping rates. \textbf{(a)} The photon emission rate $I_L$ shows an increase and saturation above $N\Gamma$ for sufficient pumping rates. \textbf{(b)} The spectral linewidth $\Delta \nu_L$ shows a decrease and eventual saturation above $\Gamma+\Gamma'$. \textbf{(c)} The spectral peak position $\omega_L$ approaches the natural emitter frequency $\omega_0$ for increasing $N$. Parameters: $N_p=N/2$, $k\delta a=\pi/5$, $\Gamma'=3\Gamma$.}
  \label{fig4}
\end{figure}
%%%%%%%%%%%%%%%%%%%%%%%%%%%%%%%%%%%%%%%%%%%%%%%%%%%%%%%%%%%%%%%%%%%%%%%%%%%%%%%%%%%%%%%%%%

%%%%%%%%%%%%%%%%%%%%%%%%%%%%%%%%%%%%%%%%%%%%%%%%%%%%%%%%%%%%%%%%%%%%%%%%%%%%%%%%%%%%%%%%%%
\begin{figure}[t]
  \centering
  \vspace{-7pt}
  \includegraphics[width=0.87\columnwidth]{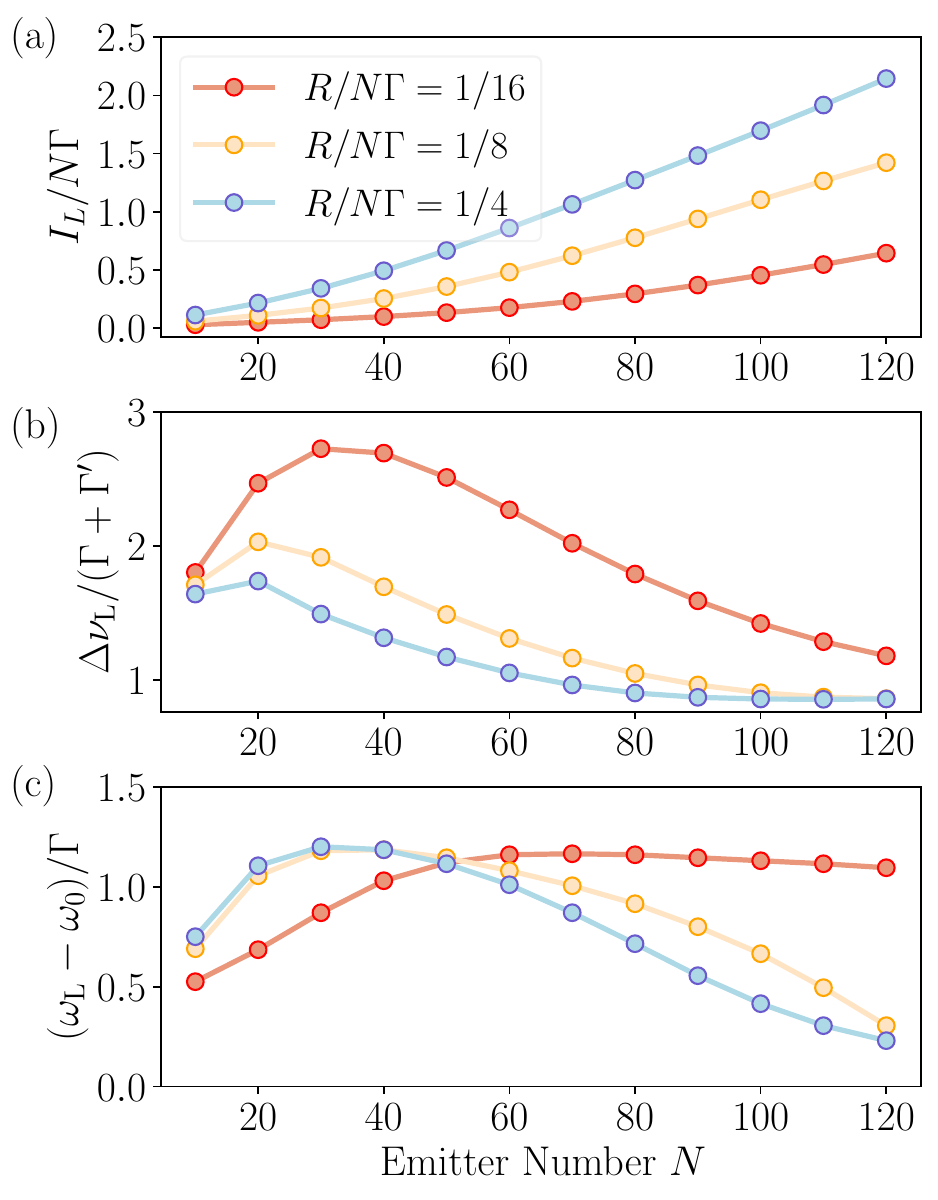}
 \caption{\textbf{Optimal emitter spacing.} Emission characteristics are shown for the left-propagating field as a function of the total emitter number $N$ for $N$-dependent pumping rates. \textbf{(a)} The steady-state emission rate $I_\mathrm{L}$ exhibits a quadratic scaling $I_L \sim \Gamma N^2$. \textbf{(b)} The full width at half maximum $\Delta \nu_\mathrm{L}$ of the spectral peak in units of $\Gamma+\Gamma'$ shows a decrease and eventually decreases below the single-emitter linewidth $\Gamma+\Gamma'$. \textbf{(c)} The spectral peak position $\omega_L$ approaches the natural emitter frequency $\omega_0$ for increasing $N$. Parameters: $N_p=N/2$, $k\delta a=\pi/N$, $\Gamma'=3\Gamma$.}
  \label{fig5}
\end{figure}
%%%%%%%%%%%%%%%%%%%%%%%%%%%%%%%%%%%%%%%%%%%%%%%%%%%%%%%%%%%%%%%%%%%%%%%%%%%%%%%%%%%%%%%%%%

The same partial-pumping geometry also produces a pronounced left--right asymmetry in the emitted field. Although the waveguide reservoir itself is reciprocal and bidirectional, the partial pumping pattern is not invariant under inversion of the chain. In combination with the propagation phases in the directional output operators, this produces different left- and right-propagating steady-state fields. As shown in Fig.~\ref{fig3}, photon emission is strongly enhanced in the left-propagating field and suppressed in the opposite direction. This asymmetry arises from the combination of spatially selective partial pumping and coherent waveguide-mediated exchange~\cite{Kasper2025emergence}. The coherent couplings \(J_{nm}\) create imaginary, phase-dependent inter-emitter coherences, which enter the directional intensities with opposite propagation phases. Consequently, the collectively emitted field interferes constructively in one propagation direction and destructively in the other.
Importantly, the bright left-propagating field is also the spectrally useful one: its emission peak remains close to the bare emitter frequency $\omega_0$, whereas the weak right-propagating component is strongly frequency shifted. We therefore focus on the left-propagating output in the following, since it simultaneously combines the desired emission properties.

Next, we fix the residual phase $k\delta a \!=\! \pi/5$ and show the scaling of emission properties with the total emitter number in Fig.~\ref{fig4}. For sufficiently strong pumping, the steady-state photon emission $I_L$ increases with $N$ and reaches values above the $N$ independent-emitter level, demonstrating superradiant enhancement. At the same time, the linewidth decreases with increasing $N$ and saturates on a scale set by the total single-emitter linewidth $\Gamma+\Gamma'$. The spectral peak position also moves toward the bare transition frequency. Thus, even for a fixed waveguide phase, increasing the number of emitters improves the three relevant properties of the emitted light: brightness, linewidth, and reduced shift of the spectral peak position. The saturation visible at large $N$, however, indicates that a fixed residual phase and pumping rate do not maintain the optimal feedback condition as the total emitter number changes.

In Fig.~\ref{fig5}, we instead consider the optimized configuration, where the residual emitter spacing is chosen according to $k\delta a=\pi/N$ and the pumping rate is scaled with $N$. In this case the quarter-wave phase relation between the pumped and passive halves is maintained as $N$ increases. The resulting emission shows the characteristic signatures of steady-state superradiance: the photon emission scales quadratically, $I_L\sim  N^2\Gamma$, the linewidth decreases toward values below the total single-emitter linewidth $\Gamma+\Gamma'$, and the spectral peak approaches the bare emitter resonance frequency $\omega_0$. 
%The optimized emitter spacing therefore realizes the desired regime of bright, narrow-line, weakly shifted collective emission.

%%%%%%%%%%%%%%%%%%%%%%%%%%%%%%%%%%%%%%%%%%%%%%%%%%%%%%%%%%%%%%%%%%%%%%%%%%%%%%%%%%%%%%%%%%
\begin{figure}[t!]
  \centering
  \includegraphics[width=0.85\columnwidth]{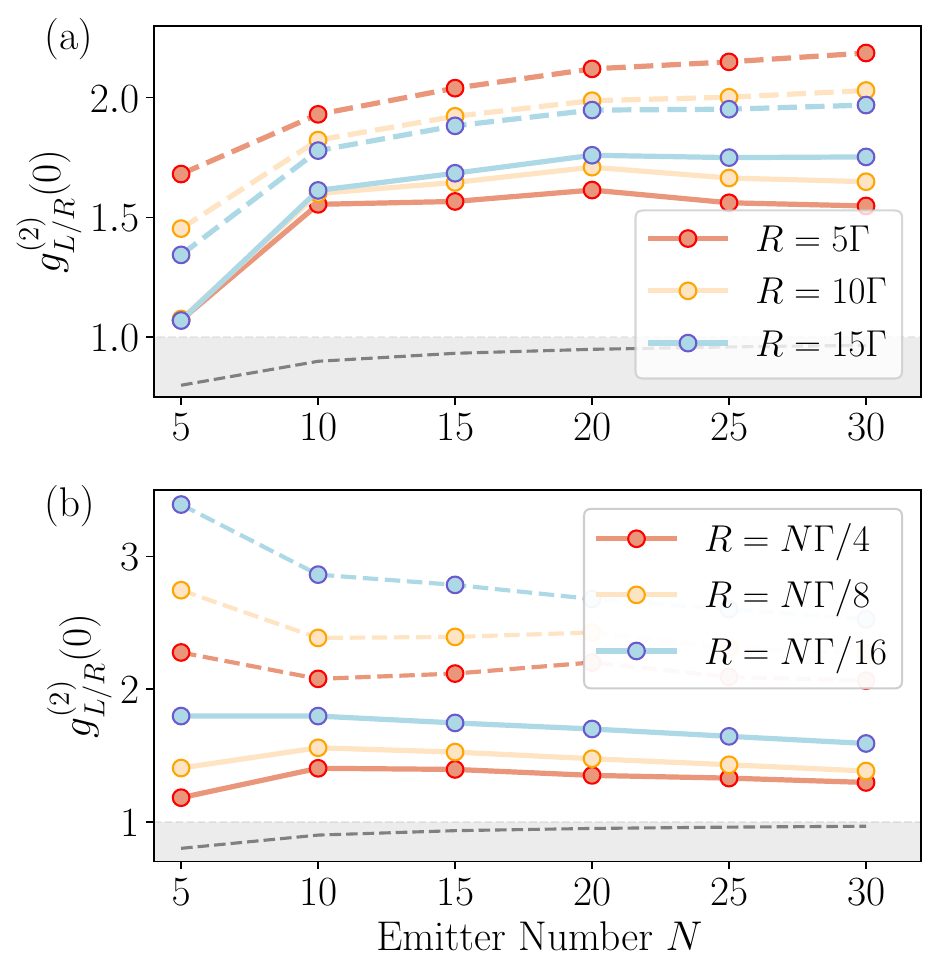}
 \caption{\textbf{Second-order correlation.} $g^{(2)}(0)$ of the left (solid line) and right (dashed line)  propagating fields as a function of $N$. \textbf{(a)} For an equidistant chain with residual phase $k\delta a=\pi/5$ and $R/\Gamma=5,10,15$. \textbf{(b)} For $k\delta a=\pi/N$ according to Eq.~\eqref{eq:eta} and $R/N\Gamma=1/16,1/8,1/4$. The left-propagating output approaches the Poissonian value from above for $k\delta a=\pi/N$, while the fixed-spacing geometry remains bunched over the range shown. In both cases \(g^{(2)}(0)\) stays below the thermal value \(2\), indicating reduced equal-time intensity fluctuations compared with strongly bunched light. Simulations are performed using a fourth-order cumulant expansion of the equations of motion. Gray dashed lines show $g^{(2)}(0)$ for $N$ independent emitters. Further parameters: $N_p =N/2$, $\Gamma'=0$.}
  \label{fig6}
\end{figure}
%%%%%%%%%%%%%%%%%%%%%%%%%%%%%%%%%%%%%%%%%%%%%%%%%%%%%%%%%%%%%%%%%%%%%%%%%%%%%%%%%%%%%%%%%%

Finally, Fig.~\ref{fig6} characterizes the photon statistics of the emitted field. We apply a fourth-order cumulant expansion of the equations of motion to account for the higher-order correlations entering $\langle(\hat{E}^\dagger)^2 \hat{E}^2 \rangle $ in Eq.~\eqref{eq:g2-function}. To derive and solve the resulting large systems of nonlinear differential equations, we employ an automated procedure based on a cumulant generating function~\cite{kubo1962generalized} with details of this approach to be published in future work. For the optimized configuration, within the computationally accessible system sizes, \(g^{(2)}_L(0)\) approaches the Poissonian value from above, while the fixed-spacing configuration remains more strongly bunched. The approach of $g^{(2)}_L(0)$ toward unity indicates reduced equal-time intensity fluctuations. Together with the narrow emission spectrum and superlinear intensity scaling, this behavior is consistent with the buildup of superradiance with a stationary laser-like output field rather than independent broadband fluorescence. %The partially pumped chain therefore realizes a continuous collective light source whose steady-state emission is bright, spectrally narrow, and increasingly locked to the bare emitter transition.

\subsection{Metrological performance}

A central motivation for narrow-linewidth superradiant emission is its use as an active optical frequency reference. In such a device, the emitted field should ideally combine three properties: a substantial photon flux, which reduces photon shot noise in the detected signal, a narrow linewidth, which increases the coherence time and improves the frequency resolution, and a small frequency shift from the bare atomic transition. The last property ensures that the output field remains close in frequency to the atomic reference rather than to geometry- or interaction-induced collective modes~\cite{Meiser2009Prospects,Kazakov2022UltimateStability,RileyHowe2008FrequencyStability}.

To quantify these requirements in a single parameter, we introduce the metrological parameter
\begin{equation} \label{eq:metro}
    M_{L/R} \equiv
    \frac{I_{L/R} \Gamma}
    {\Delta\nu_{L/R}^2+4(\omega_{L/R}-\omega_0)^2}.
\end{equation}
Here, \(I_{L/R}\) is the field intensity emitted into the left- or right-propagating mode, \(\Delta\nu_{L/R}\) is the full width at half maximum of the corresponding emission spectrum, and \(\omega_{L/R}-\omega_0\) quantifies the frequency shift away from the bare emitter resonance. The additional factor of \(\Gamma\) makes \(M_{L/R}\) dimensionless, since \(I_{L/R}\), \(\Gamma\), \(\Delta\nu_{L/R}\), and \(\omega_{L/R}-\omega_0\) are all expressed as rates or frequencies.
The structure of Eq.~\eqref{eq:metro} is motivated by shot-noise-limited frequency readout~\cite{RileyHowe2008FrequencyStability,Kazakov2022UltimateStability}. Here, the frequency uncertainty scales approximately as the spectral linewidth divided by the square root of the detected photon number, i.e. brighter emission and narrower linewidth directly improve the short-term frequency stability.
In addition, a frequency reference should be accurate: the spectral maximum should remain close to the atomic transition frequency \(\omega_0\). The term \(4(\omega_{L/R}-\omega_0)^2\) in Eq.~\eqref{eq:metro} therefore penalizes frequency shifts. Large values of \(M_{L/R}\) thus identify emission that is simultaneously bright, spectrally narrow, and weakly shifted from the atomic reference frequency.

\begin{figure}[t!]
  \centering
  \vspace{0.2em}
\includegraphics[width=0.97\columnwidth]{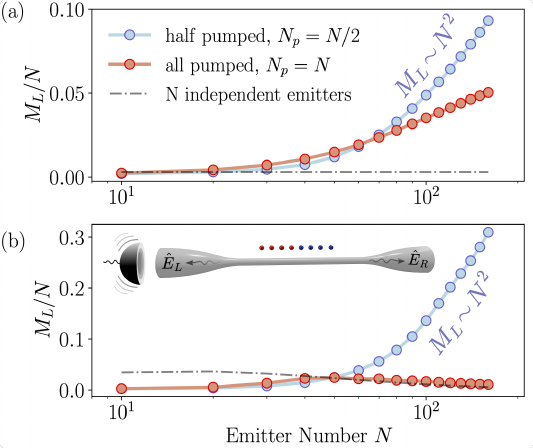}
 \caption{\textbf{Metrological performance.} Comparison of the metrological parameter $M_L$ in Eq.~\eqref{eq:metro} in the left waveguide output field. We consider partial pumping (leftmost $N_p\!=\! N/2$ emitters) and all emitters pumped ($N_p\!=\!N$) as a function of $N$. \textbf{(a)} Equidistant chain with residual phase $k\delta a=\pi/5$ and $R\!=\!15\Gamma$. \textbf{(b)} Optimal residual spacing $k\delta a=\pi/N$ and pump rate $R\!=\!N\Gamma/16$. The gray dashed line shows $M_L$ for $N$ independent emitters. Partial pumping shows the best metrological performance according to the heuristic parameter $M_L$ and fares substantially better for optimal emitter spacing. $\Gamma'=\Gamma$ in both plots.}
  \label{fig7}
\end{figure}

As a benchmark, we compare the collective emission to \(N\) independent incoherently pumped emitters. For a single independent emitter, the steady-state photon flux into one propagation direction is
\begin{equation}
    I_{L/R}^{\rm ind}
    =
    \frac{1}{2} I_{\rm ind}
    =
    \frac{1}{2}
    \frac{R\Gamma}{\Gamma+\Gamma'+R},
\end{equation}
where the factor \(1/2\) reflects symmetric emission into the left- and right-propagating waveguide modes. The corresponding linewidth is power broadened by the incoherent pumping and reads $\Delta\nu_{\rm ind} = \Gamma+\Gamma'+R$. For \(N\) independent emitters, the intensity grows only linearly with \(N\), while the linewidth is unchanged. In the absence of frequency shifts, the independent-emitter metrological parameter is therefore
\begin{equation}
    M_{L/R}^{\rm ind}
    =
    N\frac{R\Gamma^2}{2(\Gamma+\Gamma'+R)^3}.
\end{equation}
In the strong-pumping limit \(R\gg \Gamma+\Gamma'\), one obtains $M_{L/R}^{\rm ind} \sim N\Gamma^2/(2R^2)$ for fixed pump rate and $M_{L/R}^{\rm ind} \sim 1/N$ for pump rates proportional to $N$, as seen in Fig.~\ref{fig7}. Increasing the pump rate in an independent-emitter ensemble thus increases the photon flux only at the cost of strong power broadening, lowering the metrological figure of merit.

%%%%%%%%%%%%%%%%%%%%%%%%%%%%%%%%%%%%%%%%%%%%%%%%%%%%%%%%%%%%%%%%%%%%%%%%%%%%%%%%%%%%%%%%%%
%%%%%%%%%%%%%%%%%%%%%%%%%%%%%%%%%%%%%%%%%%%%%%%%%%%%%%%%%%%%%%%%%%%%%%%%%%%%%%%%%%%%%%%%%%
%%%%%%%%%%%%%%%%%%%%%%%%%%%%%%%%%%%%%%%%%%%%%%%%%%%%%%%%%%%%%%%%%%%%%%%%%%%%%%%%%%%%%%%%%%
\begin{figure}[t!]
  \centering
  \vspace{-1pt}
  \includegraphics[width=1\columnwidth]{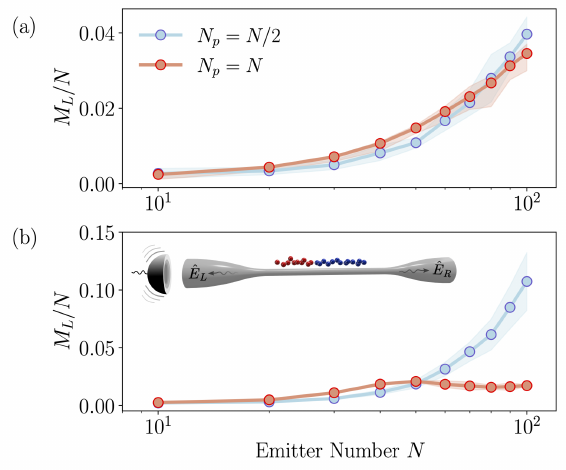}
 \caption{\textbf{Metrological performance under imperfections.} We show the disorder-averaged scaling of \(M_L\) with the total emitter number \(N\) in the presence of static positional disorder and inhomogeneous transition frequencies, modeled as described in the main text. \textbf{(a)} Pumping rate \(R\!=\!15\Gamma\). \textbf{(b)} Pumping rate \(R\!=\!N\Gamma/16\). Partial pumping shows superior metrological performance and retains an approximately quadratic scaling with \(N\). Parameters in both plots: initial residual spacing \({\delta a}=0\), \(\Gamma'=\Gamma\), disorder strengths \(\sigma_a=\lambda_0/4\), \(\sigma_\omega=\Gamma\), averaged over 50 random realizations for each curve. }
  \label{fig8}
\end{figure}
%%%%%%%%%%%%%%%%%%%%%%%%%%%%%%%%%%%%%%%%%%%%%%%%%%%%%%%%%%%%%%%%%%%%%%%%%%%%%%%%%%%%%%%%%%

The partially pumped ensemble with waveguide-mediated interactions behaves qualitatively differently. In the collective regime, the left-propagating photon flux is enhanced superlinearly, and for the optimized configuration approaches the superradiant scaling \(I_L\propto N^2\Gamma\). At the same time, the linewidth remains on the order of, or below, the single-emitter decay scale, and the spectral maximum stays close to the bare transition, \(\Delta\nu_L,\ |\omega_L-\omega_0|\lesssim \Gamma\). Consequently, Eq.~\eqref{eq:metro} yields a quadratic enhancement,
\begin{equation}
    M_L \sim N^2,
\end{equation}
as shown in Fig.~\ref{fig7}. This scaling expresses the metrological advantage of the partially pumped collective system: the waveguide-mediated interactions generate a bright output field without the severe power broadening that would occur for independent emitters and without the frequency shift that appears when all collectively interacting emitters are pumped. The unpumped part of the emitter chain therefore acts as a collective frequency-selective resonator, enabling narrow-line photon emission that remains tied to the atomic transition frequency, reminiscent of a conventional superradiant laser model~\cite{Meiser2009Prospects,Bohnet2012SteadyState}.

In Fig.~\ref{fig8}, we test the robustness of this metrological enhancement by including static positional and frequency disorder~\cite{Kasper2025emergence}. For an equidistant chain we write the emitter positions as $x_n = n(a_0+{\delta a})+\xi_n$, where $\xi_n$ are random displacements drawn from a normal distribution with standard deviation $\sigma_a$. The positional disorder \(\xi_n\) therefore randomizes the residual propagation phases entering the waveguide-mediated couplings. Inhomogeneous transition frequencies are given by $\omega_n=\omega_0+\Delta_n$ with standard deviation $\sigma_\omega$. In the frame rotating at the mean transition frequency \(\omega_0\), the additional Hamiltonian reads $\hat H_\Delta/\hbar = \sum_{n=1}^N \Delta_n \hat\sigma_n^\dagger\hat\sigma_n$. This term produces the detuning contribution
\(i(\Delta_n-\Delta_m)\langle \hat\sigma_n^\dagger\hat\sigma_m\rangle\)
in the equations of motion for the optical coherences, see the Supplemental Material for details. For each value of \(N\), the metrological figure of merit is averaged over independent disorder realizations.

The disorder-averaged results show that the metrological advantage of partial pumping persists beyond ideal emitter chains. Even for strong phase disorder, with \(\sigma_a=\lambda_0/4\) corresponding to a waveguide phase uncertainty \(k\sigma_a=\pi/2\), and in the presence of inhomogeneous transition frequencies with $\sigma_{\omega}=\Gamma$, the partially pumped configuration retains \(M_L\sim N^2\) and outperforms uniformly pumped chains. This indicates that the underlying gain--resonator mechanism with the active-passive ensemble structure holds even in the presence of disorder.
%: disorder randomizes the microscopic propagation phases, but the unpumped sub-ensemble still provides collective frequency-selective feedback for bright, narrow-linewidth emission close to the bare atomic resonance.

\section{Conclusions}

As a key result, our study shows that the presence of unpumped emitters in an ensemble with all-to-all waveguide-mediated interactions not only strongly enhances collective emission into the fundamental guided mode, but also narrows the spectral linewidth while reducing frequency shifts.
The observed improvement in the emitted light properties results from the dipole-dipole coupling of two sub-ensembles of comparable size: the incoherently pumped ensemble allows for the transfer of excitation and gain, while the unpumped sub-ensemble amplifies the emission and provides collective frequency-selective feedback. % for bright, narrow-linewidth emission close to the bare atomic resonance.
At the example of a 1D chain of quantum emitters, we show that the maximized ratio of coherent dipole-dipole exchange versus dissipative coupling between the pumped and unpumped sub-ensembles provides the optimal emitter configuration that realizes the desired regime of superradiant, narrow-linewidth emission close to the bare atomic resonance. The emitted light exhibits a strong asymmetry of emission properties in the pumped-emitter direction, and features reduced equal-time intensity fluctuations with $g^{(2)}(0)\simeq 1$. In contrast to conventional superradiant lasing in optical cavities, where the superradiant lasing threshold occurs only at large emitter numbers~\cite{Meiser2009Prospects,Kazakov2022UltimateStability}, our results point to continuous narrowband superradiance even for small numbers of emitters.
This setup thus appears to be a promising candidate for a waveguide-based active optical frequency reference with small clock-atom ensembles.
%As a key result, our study shows that waveguide-mediated long-range dipole-dipole coupling of a 1D chain of quantum emitters not only strongly enhances the coherently emitted optical power into the fundamental guided mode, but also narrows the spectral linewidth while reducing frequency shifts.
%when compared to a free-space ensemble under partial incoherent pumping~\cite{Bychek2025nanoscale}. 
%This setup thus appears to be a promising candidate for an ultrastable, accurate, and coherent frequency reference.
%At the heart of this superior performance, we identify the dipole-dipole coupling of two sufficiently large ensembles of comparable size: one pumped ensemble providing gain and a second, predominantly ground-state ensemble forming the reference oscillator. 

For instance, based on existing photonic microcell technology~\cite{Wang2019,jin2014robust}, the system should have greatly reduced technical complexity when compared to passive optical clock designs or cavity-based superradiant lasers. 
Due to the periodicity of the emitter couplings to a waveguide mode the requirement on tight atomic positioning needed for strong dipole-dipole interactions in free space can be lifted, such that the atoms no longer need to be trapped at sub-wavelength distances. Thus, emitters can be arbitrarily placed with separations that exceed the resonant wavelength because the interaction strengths depend solely on the relative phase.
This makes waveguide systems particularly beneficial for precise atomic positioning in experiments using neutral atoms trapped near optical nanofibers with optical lattices~\cite{Vetsch2010nanofiber} and reconfigurable tweezer arrays~\cite{KaufmanNi2021TweezerReview}. %Furthermore, individual pumping strategies can be developed to establish continuous incoherent pumping of emitters, for example using multi-level pumping schemes in alkaline-earth atoms or recent advancements in optical tweezer arrays to continuously reload new excited atoms into the system while maintaining coherence~\cite{chiu2025continuous}.
%At the same time, an experimental realization will require sufficient waveguide coupling efficiency, spatially selective repumping, suppression or calibration of light shifts stemming from the pumping and trapping of atoms, and precise control over emitter positions. These requirements are demanding but realistic in view of rapid progress in waveguide QED platforms. Furthermore, our results show that the observed gain-resonator mechanism with the active-passive ensemble structure persists in the presence of spatial disorder and moderate frequency broadening.
%2-3 general sentences here about results under strong psitional disorder, advantage compared to fully pumped ensemble, and the requirements for the N2 scaling of the emission intensity. 
At the same time, an experimental realization will require sufficient waveguide coupling efficiency, spatially selective repumping, suppression or calibration of light shifts stemming from the pumping and trapping of atoms, and precise control over emitter positions. These requirements are demanding but realistic in view of rapid progress in waveguide QED platforms. Importantly, our results show that the superradiant, narrow-linewidth emission enabled by the active-passive ensemble structure persists even under strong positional disorder and moderate frequency broadening. In this regime, partial pumping continues to outperform the fully pumped configuration, demonstrating that the advantage does not rely on a perfectly ordered emitter geometry. 
%The quadratic scaling of the directional emission intensity with emitter number is obtained when the pumped and unpumped sub-ensembles remain sufficiently large and comparable in size, and when their collective waveguide phases favour strong coherent exchange while suppressing net dissipative coupling between them.
%The periodic nature of emitter-waveguide couplings allows for the development of an effective scalable model for large emitter ensembles and will be studied in the future work.
%In future research, it will be interesting to exploit the periodic nature of the emitter-waveguide couplings to develop a scalable model for large emitter ensembles.
%Finally, our results show that the observed gain--resonator mechanism with an active-passive ensemble structure persists in the presence of spatial disorder and frequency broadening. Importantly, partial pumping yields superior metrological performance compared to pumping all emitters, as quantified by the metrological parameter in Eq.~\eqref{eq:metro}. It exhibits a quadratic increase with the number of emitters only under partial pumping, which persists even under complete static positional disorder. 
%A natural next step is to study emitter configurations with increased spatial symmetry~\cite{shammah2018open,lee2025exact},
 %providing a minimal analytic picture of the gain--resonator mechanism identified here. At the same time, it would allow studying the superradiant laser model~\cite{Meiser2009Prospects,Meiser2010SteadyState} beyond the fully permutation-invariant subspace in a controlled and numerically efficient way.

 \vspace{35pt}
 
\section*{Acknowledgments}
This research was funded in whole or in part by the Austrian Science Fund (FWF) QuantA 10.55776/COE1 (A.B., R.H., H.R.), and 10.55776/ESP3246525 (A.B.). M.F. acknowledges funding from the Austrian Science Fund (FWF) Grant DOI 10.55776/W1259 and from the FET OPEN Network Cryst3 funded by the European Union (EU) via Horizon 2020. S.F.Y. would like to acknowledge NSF via the CUA PFC (PHY-2317134) and AFOSR through FA9550-24-1-0311. I.V. and K.H. acknowledge support from DFG through the Collaborative Research Center SFB1227 (DQ-mat Project-ID 274200144) and from the Federal Ministry for Research, Technology and Space (BMFTR) Germany through project ATIQ.

\bibliographystyle{apsrev4-2}
\bibliography{refs}

\newpage

\clearpage
\pagebreak
\onecolumngrid
\setcounter{section}{0}

\section*{Supplemental Material}
\setcounter{equation}{0}
\setcounter{figure}{0}
\setcounter{table}{0}
%
%\makeatletter
%\renewcommand{\fnum@figure}{FIG.~S\thefigure}

%\makeatother
%
\renewcommand{\thefigure}{S\arabic{figure}}
\renewcommand{\thesection}{S\arabic{section}}
\renewcommand{\theequation}{S\arabic{equation}}

\renewcommand{\theHfigure}{S\arabic{figure}}
\renewcommand{\theHsection}{S\arabic{section}}
\renewcommand{\theHequation}{S\arabic{equation}}

\section{Quantum master equation}

We model $N$ identical two-level emitters at positions $\{x_n\}$ coupled to a single guided mode of a bidirectional one-dimensional waveguide. In the laboratory frame, the total Hamiltonian in frequency units is~\cite{Lalumiere2013InputOutput,RMP_waveguideQED_2023}
\begin{equation}
\hat H/\hbar = \sum_{n=1}^N \omega_0 \hat \sigma_n^\dagger \hat \sigma_n
+ \sum_{\mu=L,R}\int d\omega\, \omega\, \hat b_\mu^\dagger(\omega)\hat b_\mu(\omega)
+ \sum_{n,\mu}\int d\omega \left[g_\omega e^{i s_\mu k_\omega x_n}\hat b_\mu(\omega)\hat \sigma_n^\dagger + \text{H.c.}\right],
\end{equation}
where $\hat \sigma_n=|g\rangle_n\langle e|_n$, $\hat b_\mu(\omega)$ annihilates a left- or right-propagating waveguide photon, and $s_R=+1$, $s_L=-1$.

Tracing out the waveguide in the Born--Markov and rotating-wave approximations, assuming vacuum input, a linearized dispersion relation around $\omega_0$, and negligible retardation across the array, yields a Markovian master equation for the reduced emitter density matrix. In a frame rotating at $\omega_0$, it takes the form
\begin{equation}
\dot{\rho}
=
-\frac{i}{\hbar}\left[\hat H_{\rm wg},\rho\right]
+
\mathcal L_\Gamma[\rho]
+
\mathcal L_{\rm R}[\rho]
+
\mathcal L_{\rm fs}[\rho].
\end{equation}
The waveguide-mediated coherent exchange is
\begin{equation}
\hat H_{\rm wg}/\hbar=\sum_{n,m=1}^N J_{nm}\hat \sigma_n^\dagger \hat \sigma_m,
\end{equation}
while collective dissipation into the guided mode is described by
\begin{equation}
\mathcal L_\Gamma[\rho]
=
\sum_{n,m=1}^N
\frac{\Gamma_{nm}}{2}
\left(
2\hat \sigma_n\rho \hat \sigma_m^\dagger
-
\hat \sigma_n^\dagger \hat \sigma_m \rho
-
\rho \hat \sigma_n^\dagger \hat \sigma_m
\right).
\end{equation}
For a bidirectional, nonchiral waveguide one obtains
\begin{equation}
J_{nm}-\frac{i}{2}\Gamma_{nm}
=
-\frac{i\Gamma}{2}e^{ik|x_n-x_m|}.
\end{equation}
The absolute value in the phase reflects the fact that photons can propagate in both directions with equal amplitude~\cite{novotny2012principles}. In addition to the guided-mode dynamics, we include local incoherent pumping and parasitic free-space decay as independent Markovian channels,
\begin{align}
\mathcal L_{\rm R}[\rho]
&=
\sum_{n\in N_p}
\frac{R}{2}
\left(
2\hat \sigma_n^\dagger \rho \hat \sigma_n
-
\hat \sigma_n \hat \sigma_n^\dagger \rho
-
\rho \hat \sigma_n \hat \sigma_n^\dagger
\right),\\
\mathcal L_{\rm fs}[\rho]
&=
\sum_{n=1}^N
\frac{\Gamma_n'}{2}
\left(
2\hat \sigma_n \rho \hat \sigma_n^\dagger
-
\hat \sigma_n^\dagger \hat \sigma_n \rho
-
\rho \hat \sigma_n^\dagger \hat \sigma_n
\right).
\end{align}
Here, $\mathcal L_{\rm R}$ models spatially selective incoherent repumping, while $\mathcal L_{\rm fs}$ accounts phenomenologically for radiative loss outside the guided mode. Throughout this work, we assume weak frequency dependence of the waveguide coupling around $\omega_0$, negligible retardation on the scale of the emitter dynamics, and emitter separations large enough that direct free-space dipole-dipole interactions can be neglected.

%%%%%%%%%%%%%%%%%%%%%%%%%%%%%%%%%%%%%%%%%%%%%%%%%%%%%%%%%%%%%%%%%%%%%%%%%%%%%%%%%%%%%%%%%%%%%%%
\section{Waveguide phase optimization}
\label{app:phase_engineering_disorder}

In the ideal bidirectional waveguide model, the collective couplings depend only on phase differences along the waveguide, rather than on bare distances. For an equidistant chain with positions $x_n=(n-1)a$, we write the physical nearest-neighbor spacing as
\begin{equation}
    a=a_0+\delta a,\qquad a_0=\ell\lambda_0,\qquad \ell\in\mathbb{N}.
\end{equation}
The integer-wavelength part $a_0$ can be chosen large enough to suppress direct free-space dipole-dipole interactions, while the residual spacing $\delta a$ fixes the relevant waveguide phase
\begin{equation}
    \phi \equiv k\delta a = ka \quad (\mathrm{mod}\ 2\pi).
\end{equation}
The waveguide-mediated couplings then read
\begin{equation}
J_{nm}=\frac{\Gamma}{2}\sin\!\big(|n-m|\phi\big),\qquad
\Gamma_{nm}=\Gamma\cos\!\big(|n-m|\phi\big).
\end{equation}

To understand the optimal emitter spacing under partial pumping, consider a contiguous pumped ensemble $A=\{1,\dots,N_p\}$ followed by an unpumped ensemble $B=\{N_p+1,\dots,N\}$. The collective coupling between the two parts is governed by the complex phase sum
\begin{equation}
K_{AB}(\phi)\equiv \sum_{n\in A}\sum_{j\in B} e^{i\phi(j-n)} .
\end{equation}
Its real and imaginary parts determine the net dissipative and coherent inter-ensemble couplings,
\begin{equation}
\Sigma_{\Gamma}^{AB}=\Gamma\,\mathrm{Re}\,K_{AB},\qquad
\Sigma_{J}^{AB}=\frac{\Gamma}{2}\,\mathrm{Im}\,K_{AB}.
\end{equation}
For an equidistant chain this sum can be evaluated exactly as
\begin{equation}
K_{AB}(\phi)=
e^{iN\phi/2}\,
\frac{\sin(N_p\phi/2)\,\sin[(N-N_p)\phi/2]}{\sin^2(\phi/2)}.
\label{eq:KAB_general}
\end{equation}

The half-pumped case, $N_p=N/2$, is particularly transparent. Equation~\eqref{eq:KAB_general} reduces to
\begin{equation}
K_{AB}(\phi)=
e^{iN\phi/2}
\left[\frac{\sin(N\phi/4)}{\sin(\phi/2)}\right]^2 .
\end{equation}
Hence the net dissipative coupling between the pumped and passive halves vanishes whenever $\cos\!\left({N\phi}/{2}\right)=0$, that is $\phi={(2r+1)\pi}/{N}$ for $r=0,1,2,\dots.$

The smallest positive solution $\phi={\pi}/{N}$ corresponds to purely coherent coupling between the pumped and passive halves while keeping the coherent exchange large. In terms of the residual spacing this gives
\begin{equation}
    \delta a_{\rm opt}=\frac{\phi_{\rm opt}}{k}
    =
    \frac{\lambda_0}{2N}.
\end{equation}
Thus the corresponding physical spacing may be chosen as
\begin{equation}
    a_{\rm opt}=a_0+\delta a_{\rm opt}
    =
    \ell\lambda_0+\frac{\lambda_0}{2N},
\end{equation}
with $\ell$ large enough to suppress direct free-space interactions. For other contiguous pumped fractions, the same ensemble-sum analysis applies, although the precise optimum depends on $f=N_p/N$, the optimal nearest-neighbor phase remains of order $\phi_{\rm opt}\sim 1/N$.

This should be contrasted with the uniformly pumped chain. When all emitters are pumped, there is no active-passive ensemble structure, and the natural optimum is the fully phase-matched Dicke-like limit $\phi=0$ modulo $2\pi$, for which $J_{nm}=0$ and $\Gamma_{nm}=\Gamma$ for all emitter pairs. In that case all emitters radiate into the same bright collective mode.

The same geometric idea extends to non-equidistant chains. For arbitrary emitter positions, the relevant quantity is still the inter-ensemble phase sum
\begin{equation}
K_{AB}=\sum_{n\in A}\sum_{j\in B} e^{ik(x_j-x_n)}.
\end{equation}
The analogue of the equidistant optimum is therefore not a specific spacing, but the condition
\begin{equation}
\mathrm{Re}\,K_{AB}\approx 0,\qquad |\mathrm{Im}\,K_{AB}| \ \text{large},
\end{equation}
i.e. suppressed net dissipative coupling together with strong coherent exchange between pumped and passive emitters. For two ensembles of spatially disordered emitters centered around $\bar x_A$ and $\bar x_B$, one may write approximately
\begin{equation}
K_{AB}\approx N_A N_B\,e^{ik(\bar x_B-\bar x_A)}F_A^*(k)F_B(k),
\end{equation}
where
\begin{equation}
F_A(k)=\frac{1}{N_A}\sum_{n\in A} e^{ik(x_n-\bar x_A)},\qquad
F_B(k)=\frac{1}{N_B}\sum_{j\in B} e^{ik(x_j-\bar x_B)}
\end{equation}
are structural form factors describing the internal phase matching within each cloud. If $|F_A|$ and $|F_B|$ are close to unity, the same mechanism as above is recovered: a quarter-wave phase difference between the cloud centers,
\begin{equation}
k(\bar x_B-\bar x_A)\approx \frac{\pi}{2}\ (\mathrm{mod}\ \pi),
\end{equation}
suppresses the net dissipative inter-cloud coupling while maximizing the coherent exchange.

\section{Equations of motion and observables} \label{supplement:steady}

\subsection{Second-order cumulant expansion}

Using the full quantum master equation to describe the system dynamics leads to a number of coupled equations that grows exponentially with the atom number, which restricts exact numerical simulations to relatively small ensembles. To access larger emitter numbers, we employ a cumulant expansion method~\cite{kubo1962generalized,Kusmierek2023HigherOrderMeanField,rubies2023characterizing}. In this approach, one derives equations of motion for operator expectation values from the quantum Langevin or Heisenberg equations and truncates the resulting hierarchy by factorizing higher-order correlators in terms of lower-order ones.

The emitter frequencies can be inhomogeneously broadened, leading to a decay rate $\Gamma_n$ into the waveguide mode, incoherent pumping occurs with rate $R_n$ and the emitters can decay with rate $\Gamma'_n$ into free space. In realistic setups, local Stark shifts, Doppler shifts, or other sources of disorder can lead to a distribution of transition frequencies $\omega_n = \omega_0 + \Delta_n$. As long as this inhomogeneous broadening is weak compared to the relevant collective rates, $|\Delta_n| \ll \Gamma_n, J_{nm}$, we can work in a frame rotating at the mean frequency $\omega_0$ and neglect the detunings $\Delta_n$, while retaining the resulting emitter-dependent decay rates $\Gamma_n$ that sample the frequency-dependent waveguide density of states.

Throughout, we assume that all emitters are initially prepared in the ground state, such that $\langle \hat \sigma^{ee}_n\rangle = 0$, $\langle \hat\sigma_n^\dagger \hat\sigma_m\rangle = 0$, and $\langle \hat\sigma^{ee}_n \hat\sigma^{ee}_m\rangle = 0$ at $t=0$ for all $n,m$. We then truncate the hierarchy at second order, so that only three types of correlators acquire non-zero expectation values during the evolution, namely $\langle \hat\sigma^{ee}_n\rangle$, $\langle \hat\sigma_n^\dagger \hat \sigma_m\rangle$, and $\langle \hat \sigma^{ee}_n \hat \sigma^{ee}_m\rangle$. This yields a closed set of coupled differential equations for the second-order cumulants:
\begin{equation}
\label{eq.Heisenberg}
\begin{aligned}
\frac{d}{dt}  \langle \hat \sigma^{ee}_n \rangle =& -(\Gamma_n+\Gamma'_n+R_n) \langle \hat \sigma^{ee}_n \rangle + 2\sum^N_{k \neq n} \Re \Big\{ g_{nk}\langle \hat \sigma^\dagger_k \hat \sigma_n \rangle \Big\} +R_n\\
%%%%%%%%%%%%%%%%%%%%%%%%%%%%%%%%%%%%%%%%%%%%%%%%%%%%%%%%%%%%
\frac{d}{dt}  \langle \hat \sigma^\dagger_n \hat \sigma_m \rangle =& -\left( \frac{\Gamma_n + \Gamma'_n + R_n+\Gamma_m + \Gamma'_m + R_m}{2} -i \big(\Delta_n -\Delta_m \big)\right) \langle \hat \sigma^\dagger_n \hat \sigma_m \rangle + 2\Gamma_{nm}\langle \hat \sigma^{ee}_n \hat \sigma^{ee}_m\rangle + g_{nm}\langle \hat \sigma^{ee}_m \rangle +g_{nm}^*\langle \hat \sigma^{ee}_n \rangle \\
&- \sum_{k \neq n,m}^N \Big( g_{km}^*\langle \hat \sigma^\dagger_n \hat \sigma_k \rangle \big(2\langle \hat \sigma^{ee}_m \rangle -1 \big) + g_{kn} \langle \hat \sigma^\dagger_k \hat \sigma_m \rangle \big(2\langle \hat \sigma^{ee}_n \rangle -1 \big) \Big) \\
%%%%%%%%%%%%%%%%%%%%%%%%%%%%%%%%%%%%%%%%%%%%%%%%%%%%%%%%%%%%%%%%%%%%%%
\frac{d}{dt}  \langle \hat \sigma^{ee}_n \hat \sigma^{ee}_m\rangle =& -(\Gamma_n + \Gamma'_n + R_n+\Gamma_m + \Gamma'_m + R_m)\langle \hat \sigma^{ee}_n \hat \sigma^{ee}_m \rangle +R_n\langle \hat \sigma^{ee}_m \rangle +R_m \langle \hat \sigma^{ee}_n\rangle \\
&+2\sum^N_{k \neq n,m}\Re \Big\{ g_{km} \langle \hat \sigma^{ee}_n \rangle \langle \hat \sigma^\dagger_k \hat \sigma_m \rangle + g_{kn}\langle \hat \sigma^{ee}_m\rangle \langle \hat \sigma^\dagger_k \hat \sigma_n \rangle \Big\}, \\
\end{aligned}
\end{equation}
with $n\neq m$ and where we defined the collective dipole-dipole couplings $g_{nm} = i J_{nm}- \Gamma_{nm}/2$ with
\begin{equation}
J_{nm}=\frac{\sqrt{\Gamma_n \Gamma_m}}{2}\sin(k|x_n-x_m|),
\qquad
\Gamma_{nm}=\sqrt{\Gamma_n \Gamma_m} \cos(k|x_n-x_m|).
\end{equation}

\subsection{Emission spectrum}
\label{app:emission_spectrum}

The directional emission spectrum in the main text can be written as
\begin{equation}
S_{L/R}(\omega)
=
\Gamma\,\mathrm{Re}
\left[
\sum_{n,m=1}^N
e^{\pm ik(x_m-x_n)}
\int_0^\infty d\tau\, e^{i\omega\tau}
G_{nm}(\tau)
\right],
\end{equation}
where
\begin{equation}
G_{nm}(\tau)
\equiv
\left\langle
\hat\sigma_n^\dagger(t_{\rm ss}+\tau)\hat\sigma_m(t_{\rm ss})
\right\rangle .
\end{equation}

Within the second-order cumulant approximation, these two-time correlators are evaluated using the quantum regression theorem together with the same truncation used for the equal-time cumulants. Writing
\begin{equation}
p_n^{\rm ss}=\langle \hat\sigma_n^{ee}\rangle_{\rm ss},
\qquad
C_{nm}^{\rm ss}=\langle \hat\sigma_n^\dagger \hat\sigma_m\rangle_{\rm ss},
\end{equation}
the regression equations read
\begin{equation}
\frac{d}{d\tau}G_{nm}(\tau)
=
-\frac{\Gamma_n+\Gamma_n'+R_n}{2}\,G_{nm}(\tau)
-
\sum_{k\neq n}
g_{kn}\bigl(2p_n^{\rm ss}-1\bigr)\,
G_{km}(\tau),
\label{eq:regression_spectrum}
\end{equation}
with initial condition
\begin{equation}
G_{nm}(0)=C_{nm}^{\rm ss}.
\end{equation}
Equivalently, for each fixed $m$ one may write Eq.~\eqref{eq:regression_spectrum} in matrix form as
\begin{equation}
\dot{\mathbf G}^{(m)}(\tau)=\mathbf M_{\rm ss}\mathbf G^{(m)}(\tau),
\qquad
\mathbf G^{(m)}(0)=\mathbf C^{\rm ss}_{\,\cdot m},
\end{equation}
where
\begin{equation}
(\mathbf M_{\rm ss})_{nk}
=
-\frac{\Gamma_n+\Gamma_n'+R_n}{2}\delta_{nk}
-(1-\delta_{nk})\,g_{kn}\bigl(2p_n^{\rm ss}-1\bigr).
\end{equation}
The spectrum is then obtained from the Laplace transform
\begin{equation}
\tilde{\mathbf G}^{(m)}(\omega)
\equiv
\int_0^\infty d\tau\, e^{i\omega\tau}\mathbf G^{(m)}(\tau)
=
-\bigl(\mathbf M_{\rm ss}+i\omega\mathbf 1\bigr)^{-1}\mathbf G^{(m)}(0),
\end{equation}
which yields
\begin{equation}
S_{L/R}(\omega)
=
\Gamma\,\mathrm{Re}
\left[
\sum_{n,m=1}^N
e^{\pm ik(x_m-x_n)}
\tilde G_{nm}(\omega)
\right].
\end{equation}
In practice, we first solve the steady-state second-order cumulant equations for $p_n^{\rm ss}$ and $C_{nm}^{\rm ss}$, then construct $\mathbf M_{\rm ss}$ and evaluate the above linear response for each frequency $\omega$.

As a useful benchmark, consider $N$ independent incoherently pumped emitters, for which all off-diagonal couplings vanish, $g_{nm}=0$ for $n\neq m$, and
\begin{equation}
C_{nm}^{\rm ss}=\delta_{nm}p_e,
\qquad
p_e=\frac{R}{\Gamma+\Gamma'+R}.
\end{equation}
Equation~\eqref{eq:regression_spectrum} then reduces to
\begin{equation}
G_{nm}(\tau)
=
\delta_{nm}\,p_e\,
e^{-(\Gamma+\Gamma'+R)\tau/2},
\end{equation}
so that the left- and right-going spectra are identical and Lorentzian,
\begin{equation}
S_{L/R}^{\rm ind}(\omega)
=
N\,\frac{\Gamma}{2}\,p_e\,
\frac{\Gamma+\Gamma'+R}
{\omega^2+\left[(\Gamma+\Gamma'+R)/2\right]^2}.
\end{equation}
Thus, the independent-emitter reference has full width at half maximum
\begin{equation}
\Delta\nu_{\rm ind}=\Gamma+\Gamma'+R.
\end{equation}
In the absence of free-space loss, $\Gamma'=0$, this reduces to $\Delta\nu_{\rm ind}=\Gamma+R$.

\subsection{Second-order correlation function}
\label{app:second_order_correlation}

We expand the numerator of the second-order correlation function $g^{(2)}(0)$ from the main text in terms of lower-order cumulants. Starting from the directional output field
\begin{equation}
\hat E_\mathrm{L/R}
=
i\sqrt{\frac{\Gamma}{2}}
\sum_{n=1}^N e^{\pm ikx_n}\hat\sigma_n,
\end{equation}
the four-point correlator entering $g^{(2)}(0)$ is
\begin{align}
\big\langle 
\hat E_\mathrm{L/R}^\dagger 
\hat E_\mathrm{L/R}^\dagger 
\hat E_\mathrm{L/R} 
\hat E_\mathrm{L/R} 
\big\rangle
&=
\left(\frac{\Gamma}{2}\right)^2
\sum_{n,m,o,p=1}^N
e^{\pm ik(x_o+x_p-x_n-x_m)}
\langle
\hat \sigma_n^\dagger
\hat \sigma_m^\dagger
\hat \sigma_o
\hat \sigma_p
\rangle .
\end{align}

We employ a fourth-order cumulant expansion for smaller emitter numbers to capture the effect of these four-point correlators. Since $g^{(2)}_{L/R}(0)$ probes equal-time intensity fluctuations, values approaching unity should be interpreted as reduced intensity noise. Temporal phase coherence is instead inferred from the first-order field correlation and the corresponding spectral narrowing.

\section{Emission spectrum for two emitters}\label{Spectrum_2emitters}

For two emitters the set of differential equations in second order is already exact. For such a small system we calculate the spectrum analytically. Here we consider pumping only the first atom and we investigate how the couplings modify the spectrum. It is important to note that the following calculations are done with a general $\Gamma_{12}$ and $J_{12}$; the results therefore also apply to couplings mediated by any setup (e.g. atoms in free space), not just in the case of a waveguide. In other environments apart from a waveguide, one might be able to access more of the parameter region of $\Gamma_{12}$ and $J_{12}$, whereas in the waveguide the propagation phase fixes both coupling rates. \newline 
The necessary correlation functions for obtaining the spectrum are $G_{11}(\tau)$, $G_{12}(\tau)$, $G_{21}(\tau)$ and $G_{22}(\tau)$. The necessary steady-state values can be derived from a set of equations containing $\langle \sigma_1^\mathrm{ee} \rangle$, $
\langle \sigma_2^\mathrm{ee} \rangle$, $\langle \sigma_1^\dagger\sigma_2 \rangle$, $\langle \sigma_1\sigma_2^\dagger \rangle$ and $\langle \sigma_1^\mathrm{ee}\sigma_2^\mathrm{ee} \rangle$. The Quantum-Fluctuation-Regression Theorem allows us to find equations of motion for $G_{ij}(\tau) = \sigma_i^\dagger(\tau) \sigma_j(0)$ from the equations of $\sigma^\dagger_i$:
Starting from the equations of motion for $\langle \sigma^\dagger_i \rangle$
\begin{equation}
    \frac{d}{dt} \left( \begin{matrix}
        \langle \sigma^\dagger_1(\tau) \rangle \\
        \langle \sigma^\dagger_2(\tau) \rangle \\
    \end{matrix} \right)
    = 
   \underbrace{\left( \begin{matrix}
-\frac{\Gamma +R}{2} & -(i J_{12}-\Gamma_{12}/2) \cdot \sigma_{1, ss}^\mathrm{z}  \\
-(i J_{12}-\Gamma_{12}/2) \cdot \sigma_{2, ss}^\mathrm{z} & -\frac{\Gamma}{2}  \\
\end{matrix} \right) }_{=: M} \cdot
\left( \begin{matrix}
    \langle \sigma^\dagger_1(\tau) \rangle \\
        \langle \sigma^\dagger_2(\tau) \rangle \\
\end{matrix} \right),
\end{equation}
the Quantum Fluctuation Regression Theorem allows us to insert the operator $ \sigma_j(0)$ into the expectation values, such that we end up with a system of equations 
\begin{equation} \label{eq.spectrum1}
    \frac{d}{dt} y(\tau) = M \cdot y(\tau)
\end{equation}
with $y = (G_{1j}, G_{2j})^T$. Defining the Laplace transform $x(s) = \mathcal{L}(y(\tau))$, we note that the spectrum is given by $S(\omega)= 2 \Re{x(i \omega)}$, as the Fourier transform and the Laplace transform are equivalent at $s=i\omega$. In order to find $x(i\omega)$, we use the Laplace transform to convert the set of differential equations (\ref{eq.spectrum1}) into a set of algebraic equations:
\begin{equation}
    (i\omega-M) \cdot x(i\omega) = y(0)
\end{equation}
which can be easily solved. The solution to $(i\omega-M)^{-1}:=A$ is given by
\begin{subequations}
\begin{align}
A_{11} =& \frac{2 (\Gamma +2 i \omega )}{(\Gamma +2 i \omega ) (R+\Gamma +2 i \omega )-\sigma_{1, ss}^\mathrm{z} \sigma_{2, ss}^\mathrm{z} (\Gamma_\mathrm{12}-2
i J_\mathrm{12})^2} \\
A_{12} =& \frac{2 \sigma_{1, ss}^\mathrm{z} (\Gamma_\mathrm{12}-2 i J_\mathrm{12})}{(\Gamma +2 i \omega ) (R+\Gamma +2 i \omega )-\sigma_{1, ss}^\mathrm{z} \sigma_{2, ss}^\mathrm{z} (\Gamma_\mathrm{12}-2 i J_\mathrm{12})^2} \\
A_{21} =& \frac{2 \sigma_{2, ss}^\mathrm{z}(\Gamma_\mathrm{12}-2 i J_\mathrm{12})}{(\Gamma +2 i \omega ) (R+\Gamma +2 i \omega )-\sigma_{1, ss}^\mathrm{z} \sigma_{2, ss}^\mathrm{z} (\Gamma_\mathrm{12}-2 i J_\mathrm{12})^2} \\
A_{22} =& \frac{2 (R+\Gamma +2 i \omega )}{(\Gamma +2 i \omega ) (R+\Gamma +2 i \omega )-\sigma_{1, ss}^\mathrm{z} \sigma_{2, ss}^\mathrm{z} (\Gamma_\mathrm{12}-2
i J_\mathrm{12})^2}
\end{align}
and $y(0)$ can be determined from the steady state values of the corresponding operators. In conclusion, the full spectrum is then given by $S(\omega) = 2 \Re{s(\omega)}$
\begin{align} \label{eq.fullspectrum}
s(\omega) = & A_{11} \cdot \Bigl( \Gamma \langle \sigma_1^\mathrm{ee} \rangle_\mathrm{ss} + \Gamma_{12} \langle \sigma_1^\dagger\sigma_2 \rangle_\mathrm{ss} \Bigl) +  \\ \nonumber
&A_{12} \cdot \Bigl( \Gamma \langle \sigma_1\sigma_2^\dagger \rangle_\mathrm{ss} + \Gamma_{12} \langle \sigma_2^{ee} \rangle_\mathrm{ss} \Bigl) + \\ \nonumber
&A_{21} \cdot \Bigl( \Gamma_{12} \langle \sigma_1^{ee} \rangle_\mathrm{ss} + \Gamma \langle \sigma_1^{+} \sigma_2 \rangle_\mathrm{ss} \Bigl) + \\ \nonumber
&A_{22} \cdot \Bigl( \Gamma_{12} \langle \sigma_1 \sigma_2^\dagger \rangle_\mathrm{ss}  + \Gamma \langle \sigma_2^{ee} \rangle_\mathrm{ss} \Bigl),
\end{align}
\end{subequations}
where the steady state values stem from the term $y(0)$. The only thing left to do is find analytic expressions for the steady state values: From the cumulant expansion one gets a set of differential equations in the form of
\begin{equation}
\dot{\mathbf{y}} = \tilde{M} \cdot \mathbf{y} + \mathbf{x}
\end{equation}
where
\begin{align*}
\mathbf{y} = \left( \begin{matrix}
\langle \sigma_1^\mathrm{ee} \rangle  \\
\langle \sigma_2^\mathrm{ee} \rangle \\
\langle \sigma_1^\dagger\sigma_2 \rangle \\
\langle \sigma_1\sigma_2^\dagger \rangle \\
   \langle \sigma_1^\mathrm{ee}\sigma_2^\mathrm{ee} \rangle \\
\end{matrix} \right)
&& 
\mathbf{x} = \left( \begin{matrix}
R  \\
0 \\
0 \\
0 \\
0 \\
\end{matrix} \right)
\end{align*}
In steady state, we require $\dot{\mathbf{y}}_\mathrm{ss}=0$, therefore we have for $\mathbf{y}_{ss}$
\begin{equation}
\mathbf{y}_{ss} = - \tilde{M}^{-1} \cdot \mathbf{x}
\end{equation}
with the matrix $\tilde{M}$
\begin{equation}
\tilde{M} = \left(
\begin{matrix}
 -\Gamma -R & 0 & -i \cdot J_\mathrm{12}-\Gamma_\mathrm{12}/2 & i \cdot J_\mathrm{12}-\Gamma_\mathrm{12}/2 & 0 \\
 0 & -\Gamma  & i \cdot J_\mathrm{12}-\Gamma_\mathrm{12}/2 & -i \cdot J_\mathrm{12}-\Gamma_\mathrm{12}/2 & 0 \\
 -i \cdot J_\mathrm{12}-\Gamma_\mathrm{12}/2 & i \cdot J_\mathrm{12}-\Gamma_\mathrm{12}/2 & -\Gamma -R/2 & 0 & 2 \cdot \Gamma_\mathrm{12} \\
 i \cdot J_\mathrm{12}-\Gamma_\mathrm{12}/2 & -i \cdot J_\mathrm{12}-\Gamma_\mathrm{12}/2 & 0 & -\Gamma -R/2 & 2 \cdot \Gamma_\mathrm{12} \\
 0 & R & 0 & 0 & -2\Gamma -R \\
\end{matrix}
\right).
\end{equation}
Inverting this matrix one arrives at the steady state values for the expectation values:
\begin{subequations}
\begin{align}
\langle \sigma_1^\mathrm{ee} \rangle_\mathrm{ss} &=\frac{R (R+2 \Gamma ) \left(R^2 \Gamma +4 R \Gamma ^2+4 \Gamma ^3+3 R \Gamma_\mathrm{12}^2-2 \Gamma  \Gamma_\mathrm{12}^2+4
(R+2 \Gamma ) J_\mathrm{12}^2\right)}{\Gamma  (R+\Gamma ) (R+2 \Gamma )^3+(R+2 \Gamma ) \left(3 R^2-4 \Gamma ^2\right) \Gamma_\mathrm{12}^2+4
\left((R+2 \Gamma )^3+4 (R-2 \Gamma ) \Gamma_\mathrm{12}^2\right) J_\mathrm{12}^2} \\
\langle \sigma_2^\mathrm{ee} \rangle_\mathrm{ss} &=\frac{R (R+2 \Gamma )^2 \left(\Gamma_\mathrm{12}^2+4 J_\mathrm{12}^2\right)}{\Gamma  (R+\Gamma ) (R+2 \Gamma )^3+(R+2 \Gamma ) \left(3 R^2-4 \Gamma ^2\right) \Gamma_\mathrm{12}^2+4 \left((R+2
\Gamma )^3+4 (R-2 \Gamma ) \Gamma_\mathrm{12}^2\right) J_\mathrm{12}^2} \\
\langle \sigma_1^+\sigma_2^- \rangle_\mathrm{ss} &=-\frac{R (\Gamma_\mathrm{12}+2 i J_\mathrm{12}) \left(\Gamma  (R+2 \Gamma )^2+4 i (R-2 \Gamma ) \Gamma_\mathrm{12} J_\mathrm{12}\right)}{\Gamma  (R+\Gamma ) (R+2 \Gamma )^3+(R+2 \Gamma ) \left(3 R^2-4 \Gamma ^2\right) \Gamma_\mathrm{12}^2+4 \left((R+2 \Gamma )^3+4 (R-2 \Gamma ) \Gamma_\mathrm{12}^2\right) J_\mathrm{12}^2} \\
\langle \sigma_1^-\sigma_2^+ \rangle_\mathrm{ss} &=-\frac{R (\Gamma_\mathrm{12}-2 i J_\mathrm{12}) \left(\Gamma  (R+2 \Gamma )^2-4 i (R-2 \Gamma ) \Gamma_\mathrm{12} J_\mathrm{12}\right)}{\Gamma  (R+\Gamma ) (R+2 \Gamma )^3+(R+2 \Gamma ) \left(3 R^2-4 \Gamma ^2\right) \Gamma_\mathrm{12}^2+4 \left((R+2 \Gamma )^3+4 (R-2 \Gamma ) \Gamma_\mathrm{12}^2\right) J_\mathrm{12}^2} \\
\langle \sigma_1^\mathrm{ee}\sigma_2^\mathrm{ee} \rangle_\mathrm{ss} &=\frac{R^2 (R+2 \Gamma ) \left(\Gamma_\mathrm{12}^2+4 J_\mathrm{12}^2\right)}{\Gamma  (R+\Gamma
) (R+2 \Gamma )^3+(R+2 \Gamma ) \left(3 R^2-4 \Gamma ^2\right) \Gamma_\mathrm{12}^2+4 \left((R+2 \Gamma )^3+4 (R-2 \Gamma ) \Gamma_\mathrm{12}^2\right)
J_\mathrm{12}^2}.
\end{align}
\end{subequations}
Inserting these expressions into Eq.~(\ref{eq.fullspectrum}), one arrives at an analytic expression for the spectrum depending only on the system parameters $\Gamma$, $\Gamma_{12}$, $J_{12}$ and $R$. The full expression is rather lengthy, but one can do more analytic work in some limiting cases. For $\Gamma_{12}=J_{12}=0$, one arrives at the spectrum for independent atoms.
\subsection{Narrowing of the spectrum due to $J_{12}$ ($\Gamma_{12}=0$)}
\begin{figure}[ht]
  \centering
  \includegraphics[width=0.68\linewidth]{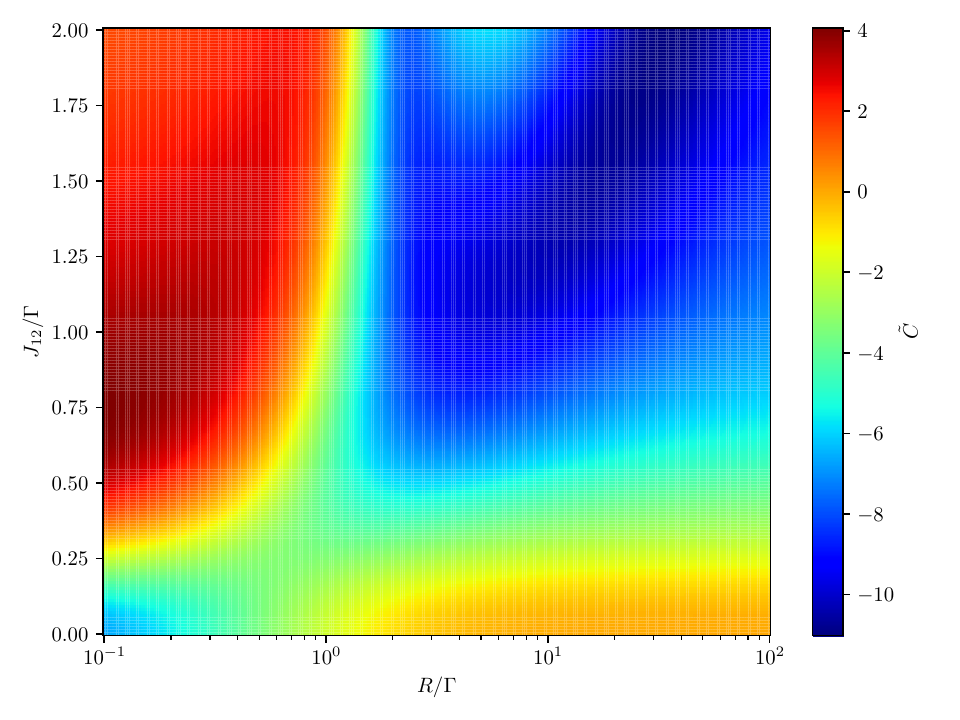}
  \vspace{-1em}
  \caption{\textbf{Scan of the normalized curvature $\tilde{C}$ at $\omega = 0$ for varying pumping strength $R$ and coupling rate $J_{12}$. All values can be calculated analytically.} Positive curvature (red) means that there is a dip in the spectrum at $\omega=0$, whereas negative curvature (blue) indicates a peak in the spectrum. For $R > \Gamma$, increasing the coupling rate $J_{12}$ leads to a narrowing of the peak. The coupling to the unpumped atom mediated by the waveguide therefore leads to a spectrum with a narrower peak.}
  \label{supp:fig_curvature_scan}
\end{figure}

\begin{figure}[t]
  \centering
  \includegraphics[width=0.92\linewidth]{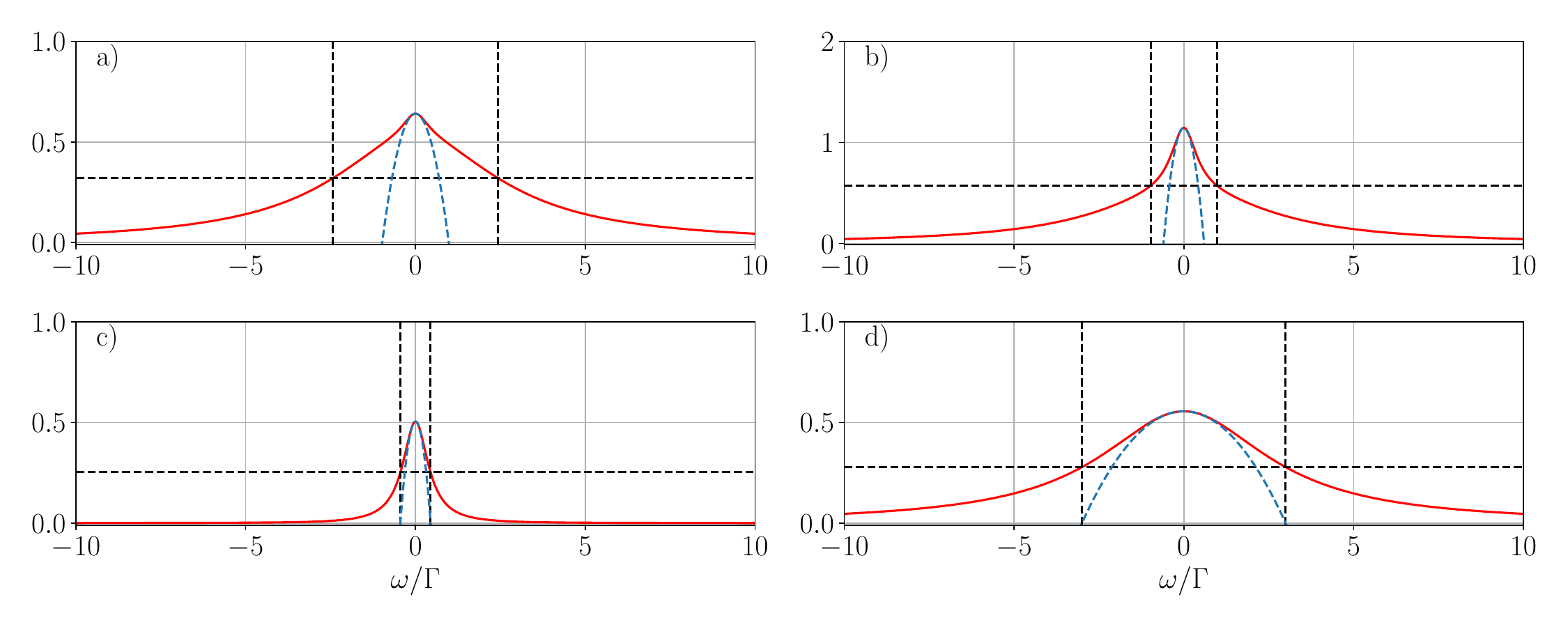}
  \vspace{-1em}
  \caption{\textbf{Analytic spectra (red), Full Width at Half Maximum (black dashed) and Taylor approximation up to second order (blue dashed) of the spectra of two emitters with propagation phase $kd=\pi/2$ modulo $\pi$. All plotted graphs are described by analytic functions.} In (a) we plot the spectrum of the pumped atom, while in (c) the spectrum of the unpumped atom is depicted. Panel (b) shows the total spectrum, while in (d) the spectrum of an independent atom is plotted. The spectrum of the pumped atom in (a) is broadened by power, whereas the spectrum of the unpumped atom in (c) is narrow. Since the total spectrum is just a sum of the individual contributions ($\Gamma_{12}=0$), we get a broader background from the pumped atom with a small peak from the unpumped atom. Comparing this total spectrum to the spectrum of a pumped atom without the coupling $J_{12}$ in (d), we clearly notice the narrowing of the peak.}
  \label{supp:fig_single_spectra}
\end{figure}

Particularly interesting is the case for $\Gamma_{12}=0$, which we analyze further. In the waveguide, this happens when the propagation phase between the emitters satisfies $kd = \pi/2 + q\pi$, or equivalently $d=(q+1/2)\lambda_0/2$ with $q\in\mathbb{N}$, then also the coupling rate $J_{12}$ is fixed to $J_{12}=\Gamma/2$. To make the following calculations more versatile, we consider a general $J_{12}$. \newline
The integer part of this distance may be chosen large enough to suppress direct free-space interactions. In this case we have no collective dissipative cross terms, so we distinguish between the spectrum of the pumped atom and the spectrum of the unpumped atom. Simplifying the above equations yields
\begin{equation} \label{eq.structure}
S_{1,2}(\omega) = 2\cdot \frac{2R\Gamma}{(R+2\Gamma)\cdot(R\Gamma +\Gamma^2+4J_{12}^2)} \cdot \frac{a_{1,2}+4\omega^2 b_{1,2}}{(4\omega^2+c)^2+d} =: K \cdot \frac{a_{1,2}+4\omega^2 b_{1,2}}{(4\omega^2+c)^2+d}
\end{equation}
with 
\begin{subequations}
\begin{align}
a_1(R, \Gamma, J_{12}) &= (\Gamma^2 (R+2\Gamma)+4\; J_{12}^2 \Gamma(1+\sigma^z_1)) \cdot (R\Gamma+\Gamma^2+4\; J_{12}^2 \sigma^z_1 \sigma^z_2) \\
b_1(R,\Gamma, J_{12}) &=  \Gamma (R+\Gamma)(R+2\Gamma)+4J_{12}^2(\Gamma(1-\sigma^z_1)+R)\\
a_2(R, \Gamma, J_{12}) &= 4 J_{12}^2 (R+\Gamma(1-\sigma^z_2))\cdot (R \Gamma+\Gamma^2+4 \; J_{12}^2 \sigma^z_1\sigma^z_2) \\
b_2(R,\Gamma, J_{12}) &= 4J_{12}^2 \Gamma (1+\sigma^z_2) \\
c(R, \Gamma, J_{12}) &= \Gamma R + \Gamma^2 - 4\; J_{12}^2 \sigma^z_1 \sigma^z_2 +R^2/2 \\
d(R, \Gamma, J_{12}) &= -\frac{1}{4} (R+2\Gamma)^2 (R^2-16 \; J_{12}^2 \sigma^z_1 \sigma^z_2),
\end{align}
\end{subequations}
where the index ${1,2}$ indicates the spectrum of the first (pumped) atom and the second (unpumped) atom. For the full spectrum, use $a=a_1+a_2$ and $b=b_1+b_2$. Approximating the spectrum by making use of the Taylor expansion up to second order
\begin{equation}
S_{1,2}(\omega) \approx S_{1,2}(0)+\frac{1}{2} \omega^2 \cdot \underbrace{K \cdot \Bigl( \frac{8\;b_{1,2}}{c^2+d}-\frac{16 \; a_{1,2} \; c}{(c^2+d)^2} \Bigl)}_{=:C} + O(\omega^4)
\end{equation}
also allows us to read off the curvature $C$ of the spectrum at $\omega=0$. Since the spectrum is symmetric, this approximation is correct up to fourth order. In Fig.~\ref{supp:fig_curvature_scan} we depict the normalized curvature $\tilde{C} = \frac{C}{S_(0)}$ for different pumping strengths $R$ and coupling rates $J_{12}$.
From Eq.~(\ref{eq.structure}) we can also determine the FWHM by starting from
\begin{equation}
\frac{a_{1,2}+4\omega^2 b_{1,2}}{(4\omega^2+c)^2+d} = \frac{a_{1,2}}{c^2+d} \frac{1}{2}.
\end{equation}
The solutions of this equation read
\begin{equation}
(\omega^2)_\pm = \frac{-ac+b(c^2+d)}{4a} \pm \frac{1}{32 a} \sqrt{64a^2(c^2+d)+(8ac-8b(c^2+d))^2}.
\end{equation}
We disregard the solution with the minus, so the solution for the FWHM is then, because of symmetry, given by
\begin{equation}
\mathrm{FWHM}=2 \cdot \sqrt{(\omega^2)_+}.
\end{equation}
Altogether we arrive at 
\begin{equation}
\mathrm{FWHM}=\sqrt{\frac{1}{a}} \cdot \sqrt{-ac+b(c^2+d)+1/8 \sqrt{64a^2(c^2+d)+(8ac-8b(c^2+d))^2}}.
\end{equation}
Armed with the analytic expression for the full spectrum, the spectra of the individual atoms, their curvature and FWHM, we now investigate the influence of the coupling mediated through the waveguide on the spectra in Fig.~\ref{supp:fig_single_spectra}. The spectrum of the pumped atom (a) gets power broadened, while the spectrum of the unpumped atom (c) is narrow. The sum of these spectra is depicted in (c). Comparing the full spectrum of both atoms (c) to the spectrum of an independent pumped atom in (d), the significant narrowing of the peak becomes apparent.
\subsection{Independent emitters ($J_{12}=0$, $\Gamma_{12}=0)$}
For independent emitters the above equations simplify further to
\begin{equation}
S(\omega) = \frac{2R\Gamma}{(R+\Gamma)^2} \cdot \frac{1}{1+(\frac{2\omega}{R+\Gamma})^2}.
\end{equation}
Here we have $\omega_\mathrm{max}=0$ and for the FWHM linewidth we get power broadening,
\begin{equation}
\Delta \omega = \Gamma + R.
\end{equation}
\end{document}